%% file: 0_main.tex
\documentclass[conference]{IEEEtran}

\usepackage{cite}
\usepackage{graphicx}
\usepackage{tikz, amsmath, amssymb, bm, color}
\usepackage{lipsum, float}

\usepackage{bbding} % tick marks for flowcharts

\tikzset{
    block/.style    = {draw, thin, rectangle, minimum height = 3em,
    minimum width = 3em}
}
\usepackage[ruled, vlined, linesnumbered]{algorithm2e}
\usepackage{mathtools}
\usepackage{siunitx}
\definecolor{niceyellow}{HTML}{fbe396}
\usepackage[disable, color=niceyellow,linecolor=gray]{todonotes}

\usepackage{enumitem, soul}
\newlist{todolist}{itemize}{2}
\setlist[todolist]{label=$\square$}
\usepackage{pifont}

\SetKw{Continue}{continue}

\begin{document}
\title{Setting the Yardstick: A Quantitative Metric for \textit{Effectively} Measuring Tactile Internet}

% \author{}
% Double blind review, hide this until publish
\author{J.P.~Verburg$^*$\textsuperscript{\textsection}, H.J.C.~Kroep$^*$\textsuperscript{\textsection}, V.~Gokhale$^*$, R.~Venkatesha~Prasad$^*$, and V.~Rao$^\dagger$\\
$^*$Delft University of Technology, The Netherlands, $^\dagger$Cognizant Technology Solutions, The Netherlands
% Email: j.p.verburg@student.tudelft.nl, h.j.c.kroep@tudelft.nl, r.r.venkateshaprasad@tudelft.nl and v.rao@tudelft.nl
}
\vspace{-10pt}
\maketitle
\begingroup\renewcommand\thefootnote{\textsection}
\footnotetext{J.P. Verburg and H.J.C. Kroep assert equal contributions.}
\endgroup
%
% \input{todo.tex}
% \clearpage

\begin{abstract}
The next frontier in communications is \textit{teleoperation} -- manipulation and control of remote environments. Compared to conventional networked applications, teleoperation poses widely different requirements, ultra-low latency (ULL) being the primary one. Teleoperation, along with a host of other applications requiring ULL communication, is termed as \textit{Tactile Internet} (TI). A significant redesign of conventional networking techniques is necessary to realize TI applications. Further, these advancements can be evaluated only when meaningful performance metrics are available. However, existing TI performance metrics fall severely short of comprehensively characterizing TI performance. In this paper, we take the first step towards bridging this gap. To this end, we propose a method that captures the fine-grained performance of TI in terms of delay and precision. We take Dynamic Time Warping (DTW) as the basis of our work and identify whether it is sufficient in characterizing TI systems. We refine DTW by developing a framework called \textit{Effective Time- and Value-Offset (ETVO)} that extracts fine-grained time and value offsets between input and output signals of TI.
%The strength of the proposed concepts, together referred to as , primarily lies in the fact that they can quantitatively characterize the performance of any end-to-end TI system in a manner that is agnostic to underlying implementation details.
Using ETVO, we present two quantitative metrics for TI -- \textit{Effective Delay-Derivative (EDD)} and \textit{Effective Root Mean Square Error}. Through rigorous experiments conducted on a realistic TI setup, we demonstrate the potential of the proposed metrics to precisely characterize TI interactions.
%compare the performances of DTW and ETVO. Our findings show that ETVO results in 62.5$\times$ less time noise for up to \SI{34}{\%} increased RMSE compared to DTW. \todo[inline]{the last sentence referring to results in no longer correct}

%We demonstrate the effectiveness (40x more accurately than DTW) of the proposed metrics and discuss its potential usage jointly with existing metrics.     

% To this end, enormous developments are being achieved in the domains of networking infrastructure, highly immersive human computer interfaces, and . However,
\end{abstract}
\input{2_introduction/introduction.tex}

\input{3_relatedwork/relatedwork.tex}

\input{4_algorithm/algorithm.tex}

\input{5_experimentalsetup/experimentalsetup.tex}
\input{6_performanceanalysis/performanceanalysis.tex}
\input{7_conclusion/conclusion.tex}

\ifCLASSOPTIONcaptionsoff
  \newpage
\fi
\bibliographystyle{IEEEtran}
\bibliography{bibs}
\end{document}

%% file: 2_introduction/introduction.tex
\section{Introduction}\label{sec:intro}
%
% Intro TI
% Good intro's are hard mkay.
%
Tactile Internet (TI)\cite{Fettweis2014} is irrefutably at the forefront amid several emerging technological innovations that are foreseen to revolutionize future industries and the lifestyle of humans. The crux of TI is its potential to enable \textit{teleoperation} - transportation of physical skills of humans for manipulation and control of remote environments. Sensory feedback, including vibrotactile, kinesthetic, audio, and video modalities, transmitted via TI, should provide a feeling of collocation between geographically distant locations. These can facilitate humans to remotely perform activities in uninhabitable or resource-constrained locations as if they are physically present there. TI is not just limited to teleoperation applications such as telesurgery, remote disaster management and telerepair of machinery, but also the other sectors like education, health, and manufacturing \cite{Aijaz2018}.

% \noindent\textbf{TI vs. the current Internet.} TI applications are significantly different from the existing real-time applications served by the Internet. For a concrete exposition,  consider telesurgery involving haptic-audio-video feedback (TI application) and real-time audio-video teleconferencing (Internet application). In the former case, the surgeon's physical actions need to be precisely transferred on the remotely located patient through a robotic teleoperator (as shown in Figure~??\todo{proper reference}). The remote environment has to be physically sensed and fed back to the surgeon instantly. The speed of the feedback determines the surgeon's perception of the remote environment and thereby his/her subsequent actions. The work in \cite{holland2019} states that for telesurgery to be realizable, a feedback latency of sub-\SI{10}{ms} is necessary. Non-compliance with the ultra-low latency (ULL) requirement results in severe impairment in the surgeon's ability to teleoperate, potentially leading to catastrophic outcomes. On the other hand, in case of teleconferencing using audio-video, latency in the order of even several hundreds of milliseconds only slows the conversation down by a small amount, however, without hampering one's ability to interact. 
% \todo{Vijay: This is not making a case for TI! Summarize arguments for TI vs Internet other than ULL.}
TI applications are significantly different from the existing real-time applications served by the Internet. The transmission of tactile sensor data and the sensory feedback have latency requirements in orders of magnitude lower than the current real-time applications on the Internet. Furthermore, some TI applications would also need ultra-high reliability, which is not yet supported on current networks. 
% To summarize, the presence of the feedback loop in TI application places extremely tight constraints on the latency requirements.
% 
\begin{figure}[t]
    \centering
    \includegraphics[width=0.8\columnwidth]{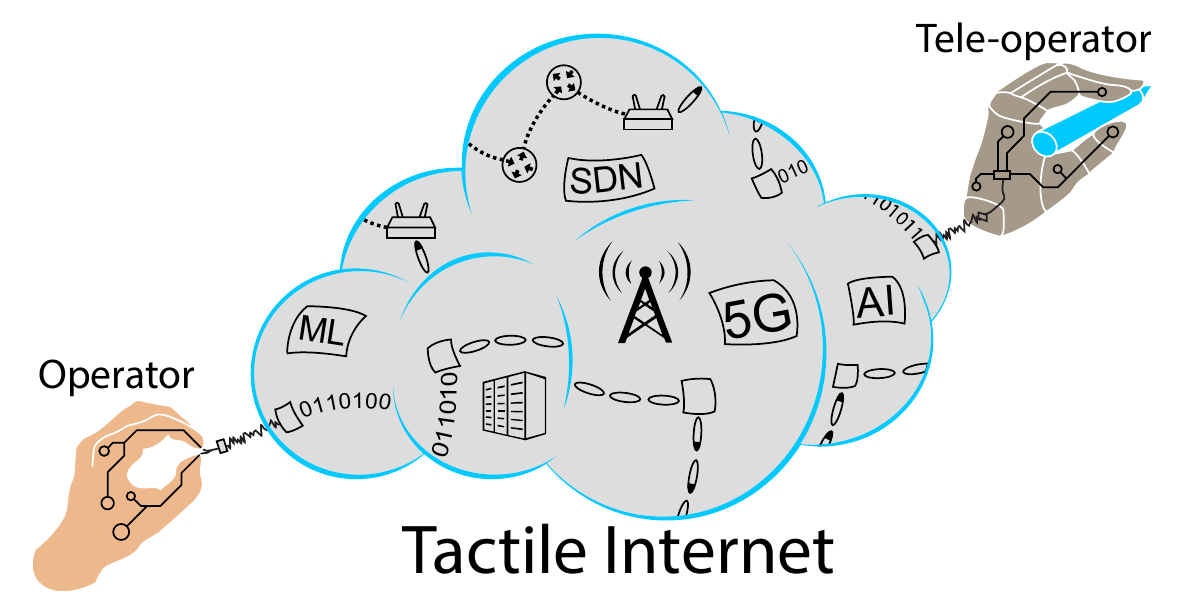}
    % \caption\hl{{A schematic representation of an E2E TI application.}}
    \caption{A schematic representation of an E2E TI application.}
    \label{fig:ti-arch}
    \vspace{-8pt}
\end{figure}
Thanks to the vision and recent advancements in the field of 5G, particularly in the context of ultra-reliable, low-latency communication (URLLC), TI will benefit enormously~\cite{Millnert2018,Samarakoon2018,Gringoli2018}.
% in addressing the ultra low-latency (ULL) constraint  
% While ULL is a requirement for TI, it is not a sufficient criterion for comparing candidate TI solutions, since there is much more to TI interactions than just latency. We provide more insights on this later in this section. With extensive research being carried out towards developing working TI solutions, it is imperative to establish metrics for measuring and comparing their performances. 

As with real-time applications on the Internet, objectively measuring the quality of a session is important for several reasons: for adapting session quality based on network dynamics,  estimating the quality of a session a priori (which is important before executing mission-critical applications), developing and benchmarking of new solutions at various layers of the protocol stack including application layer. Unlike the real-time applications on the Internet today, TI applications are complex due to the active involvement of humans in the entire control loop. However, it is imperative to establish objective metrics for measuring and comparing TI session performances. 
%This provides us with a reasonable understanding of the ability of a given TI solution to facilitate teleoperation.

\begin{figure}[!t]
    \centering
    \includegraphics[width=0.8\columnwidth]{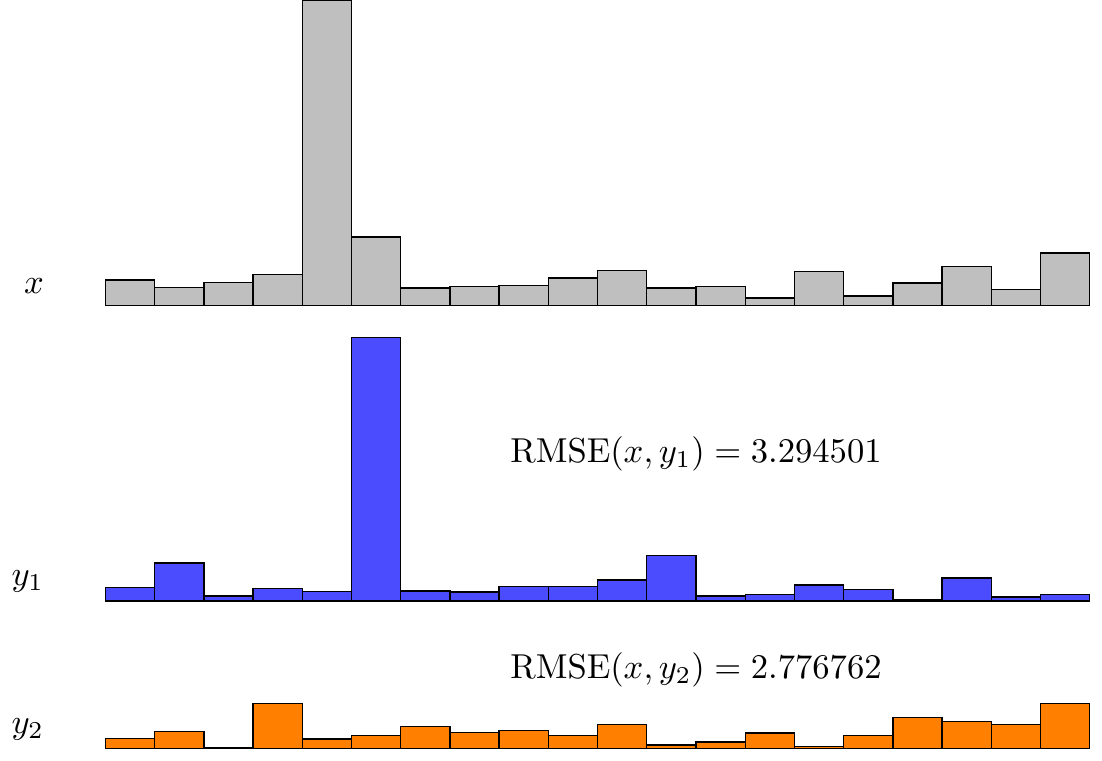}
    \caption{Demonstration of wrong conclusions drawn by RMSE although the shapes of $x$ and $y_1$  are more similar than the shapes of $x$ and $y_2$. The abscissa has units in time and the ordinate in amplitude of the signal.}
    \label{fig:ti-blackbox}
    \vspace{-10pt}
\end{figure}

\noindent\textbf{Measuring TI performance.} For fine-grained performance analysis, a TI system should be evaluated based on how quickly and accurately it can reproduce actions at the remote end. Hence, it is crucial to determine the degradation it introduces in terms of \textit{time} and \textit{value} collectively.
% As illustrated in Figure~\ref{fig:ti-arch}, TI is composed of a complex interconnection between several modules which do not necessarily reside in the network.
While Quality of Service (QoS) metrics, such as latency, reliability, and throughput, are known to characterize real-time network applications well, they fall short for TI applications. For instance, a TI solution with an average latency of \SI{15}{ms} is not necessarily inferior compared to another with \SI{10}{ms} average latency. This is because the former could be intelligently delaying the signals to avoid over-provisioning of resources when the human perception is less sensitive to latency; for example, in case of medium dynamic environments \cite{holland2019}. Similar arguments can be constructed for other QoS metrics. On the other hand, literature provides several works that propose root-mean-square error (RMSE) based Quality of Experience (QoE) metrics for qualitative performance evaluation of TI \cite{Chaudhari2011,Yuan2014}. Essentially, 
RMSE is not designed to extract time and value offsets separately. Time offset is the delay variations due to network and value offset may be due to noise or packet loss (inducing loss of data). We demonstrate this limitation in Figure~\ref{fig:ti-blackbox}. While it is easy to observe that among $y_1$ and $y_2$, the profile of $y_1$ is closer to $x$ due to the presence of the dominant peak, the RMSE of $y_1$ is \SI{19}{\%} higher than that of $y_2$. Hence, the QoE metrics do not always indicate close to real performance of TI.
%This has also been pointed out by a few existing works; see, for example, the references \cite{van2017,sharma2019}.

%Note that a combination of QoS and QoE metrics also falls short of the stated objectives.
The lack of a framework for performance characterization of TI at a fine-grained level severely impedes the overall progress of TI. This forms the primary motivation for this work.

\noindent\textbf{Our contributions.} Our approach is to devise a method that is capable of extracting time and value offset between the input and output signals of TI. The strength of this approach lies in the fact that it can be applied to any end-to-end TI system (starting from sensors on one end all the way to the actuators on the other end) in a manner that is agnostic to the TI modules. We take Dynamic Time Warping (DTW), a common method for determining sample-wise similarity between two signals, as the starting point for our work. Our contributions in this work are as follows.

\begin{itemize}
\item  We present a detailed discussion on DTW in characterizing TI sessions and show where we can further use it for TI applications. 
\item We present a concrete mathematical framework that we call \textit{Effective Time- and Value-Offset (ETVO)} that extract time and value offset between input and output signals of a TI system, which hitherto was impossible.
\item Based on ETVO, we propose two quantitative metrics -- \textit{Effective Delay-Derivative (EDD)} and \textit{Effective RMSE (ERMSE)} that can jointly characterize the TI system in a fine-grained manner.
\item Through experiments conducted on a realistic TI setup, we demonstrate the efficacy of the proposed method in supporting the characterization of a TI session close to real. 
\end{itemize}

The remainder of the paper is organized as follows. We review the literature for potential performance metrics for TI in Section \ref{sec:relatedwork}. We then provide a brief overview of DTW and its insufficiency specifically for characterization of TI interactions in Section~\ref{sec:limitations}. We present the complete design of the proposed performance metrics in Section~\ref{sec:etvo}. In Section~\ref{sec:setup}, we describe the experimental setup, and present our findings in Section~\ref{sec:results}. Finally, we state our conclusions and future directions in Section~\ref{sec:conclusions}. 

%% file: 3_relatedwork/relatedwork.tex
\section{Related Work}\label{sec:relatedwork}
%Literature offers a rich volume of works that are relevant for the metric design in this paper. 
%Our literature review spans two orthogonal domains, where one relates to the state of the art in existing performance metrics that we present in Section~\ref{subsec:performance}, and the other concerns the state of the art in techniques for detection of signal similarity which we discuss in  Section~\ref{subsec:similarity}.
\subsection{TI performance metrics}\label{subsec:performance}
%As already mentioned in Section~\ref{sec:intro}, 
Standard QoS metrics have been extensively employed, and several QoE metrics have been devised for evaluating TI systems. In this section, we present an overview of these.

\subsubsection{\textbf{QoS}}
Traditional QoS metrics for network performance include latency, jitter, reliability, and throughput.
Several modular designs of TI systems use QoS for characterizing TI performance. While \textit{Admux}, an adaptive multiplexer for TI proposed by Eid et al. \cite{Eid2011}, uses all of the above metrics, the multiplexing scheme by Cizmeci et al. focuses on throughput and latency \cite{Cizmeci2017}. Hinterseer et al. \cite{Hinterseer2006} proposed a haptic codec that focuses on reducing the application throughput. The codec is based on perceptual Deadband, where packets are not sent if the changes are below the level of human perception. The congestion control scheme by  Gokhale et al. \cite{Gokhale2017} aims to contain latency and jitter within their permissible QoS limits. Further, a string of works has emerged recently that attempt to address the ULL requirement of TI by leveraging the advancements in the field of 5G networks. The works in \cite{Li2018,sachs2018,kim2018} provide a detailed discussion on the vision and progress in this direction. While QoS metrics play a vital role in characterizing the network performance, they are insufficient for comprehensive characterization of TI, as described in Section~\ref{sec:intro}.

\subsubsection{\textbf{QoE}}
%filling out responses to a detailed questionnaire regarding the perception, 
Subjective QoE metrics aim to capture the quality of teleoperation by having the human users subjectively rate their experience. A couple of works that adopt this approach include \cite{Basdogan2000, Yuan2014}. Since this method is cumbersome and resource-intensive, objective QoE metrics that estimate the quality of teleoperation as experienced by the human controller have also been designed.
Hinterseer et al.~\cite{Hinterseer2006} exploited the idea of human perception having a logarithmic relationship with the haptic stimulus. A framework was developed to validate this by using the traditional Peak Signal-to-Noise Ratio (PSNR) of the reconstructed haptic signal as the QoE metric. Later, 
Sakr et al.~\cite{Sakr2007} introduced Haptic Perceptually Weighted Peak Signal to Noise Ratio (HPW-PSNR) that is aware of human insensitivity to Deadband of the haptic signal, the idea initially presented in \cite{Hinterseer2006}.
Chaudhuri et al. followed up on this to propose Perceptual Mean Square Error (PMSE) that maps MSE to the human perceptual domain \cite{Chaudhari2011}. Recently, Hassen et al. proposed the Haptic Structure SIMilarity (HSSIM) index to improve the objective estimation of human perception. HSSIM extracts the similarity between original and reconstructed haptic signals. All of the above metrics are based primarily on RMSE, which works by combining degradation in both time and value domains. 
%\hl{Although QoE metrics are capable of measuring the human experience, they can only perform TI characterization at a granular level.}
%Hamam et al. took an entirely different approach to analytically model haptic device properties while calculating the objective metric \cite{Hamam2008}.   
%

\subsection{Signal similarity metrics}\label{subsec:similarity}
Determining the similarity between two signals is an age-old problem in the signal processing domain and has been extensively researched due to its numerous applications, such as speech and gesture recognition~\cite{Salvador2007}. In this section, we discuss some of the techniques devised for this purpose and examine their applicability in extracting time and value errors.  
%Several signal processing methods exist in the non-TI domains that seemingly manifest potential to extract time and value errors separately.
%An important aspect of TI applications is that the input from the controller should closely match the output at the rendering side, ideally matching each sample.
Cross-correlation computes the time-offset (delay) between the two signals that maximizes their dot product \cite{Rabiner1975}. It is well known that shared networks usually manifest highly non-deterministic and time-varying characteristics. Hence, a constant delay is an incorrect choice for representing the entire TI characteristics.
Another popular method known as Dynamic Time Warping (DTW) exists for signals encountering a time-varying delay. DTW conducts an exhaustive search to achieve sample-wise matching between the two signals in a manner that minimizes the cumulative Euclidean distance \cite{Sakoe1978}. It provides an extremely useful construct in determining \textit{how similar two signals are}. Since in the case of teleoperation, input and output signals are expected to be similar, our problem is to find out \textit{how two similar signals are different}. DTW falls short of achieving this objective of TI.
% Owing to its popularity, several optimization algorithms for efficient computation of DTW exist, including a few recent ones \cite{Salvador2007, Al2009, Silva2016}.
Several follow-up works on DTW exist, with each of them attempting to outperform DTW in one or more aspects. The most widely recognized ones include Edit Distance on Real sequences (EDR) \cite{Chen2005}, Edit distance with Real Penalty (ERP) \cite{Chen2004}, and Longest Common Sub-Sequence (LCSS) \cite{Vlachos2002}. However, they manifest the inherent characteristics of DTW and hence are unsuitable for TI as will be detailed in the following sections. We discuss DTW in detail in the rest of the paper as it forms the basis of the ETVO design.

%% file: 4_algorithm/algorithm.tex
\section{DTW: Background and analysis}\label{sec:limitations}
In this section, we present the necessary mathematical background for DTW and analyze its shortcomings in addressing the stated objectives of this work. DTW constructs a \emph{warp path} that indicates a sample-wise mapping between two time-series that minimizes their cumulative Euclidean distance. %Figure~\ref{tkz:shortcommings} illustrates the alignment between the two sequences as constructed by DTW. However while minimizing the cumulative distance DTW does not take into account the effect of time delay variation over the value and value va 
% \todo[inline]{Caption of Fig 3: can't really figure out what the ``insufficient of sample-wise alignment" is from the figure. Also the first point at which it is referred to is not the context wherein we elaborate insufficiencies.}
% \todo[inline, author=Kees]{It isn't insufficient for the DTW definition. It is just that it creates behavior that doesn't match with our goals}
% \todo[inline]{Our goals such as ...?}

\subsection{Mathematical Representation}
\begin{figure}[t]
\centering
\includegraphics[width=0.499\textwidth]{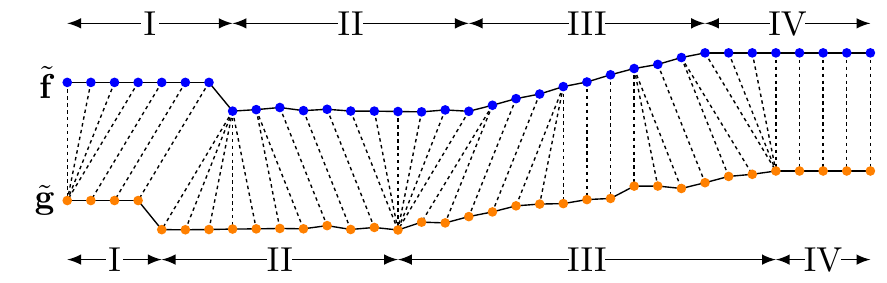}
\caption{Insufficient of sample-wise alignment of DTW. The solid lines indicate the input and output signals, and the dashed lines indicate the alignment between the samples.}
\label{tkz:shortcommings}
\vspace{-10pt}
\end{figure}

Let $\tilde{\mathbf{f}}$ and $\tilde{\mathbf{g}}$ denote the  $N$-length discrete time-series corresponding to input and output signals, respectively. Hence, $\tilde{\mathbf{f}}, \tilde{\mathbf{g}} \subset \mathbb{R}^N$. Let $\tilde{\mathbf{W}}$ denote the set of all possible warp paths between $\tilde{\mathbf{f}}$ and $\tilde{\mathbf{g}}$, where a warp path represents the sample-wise alignment. Let the $(k+1)$-th point of a warp path $\tilde{\mathbf{w}} \in \tilde{\mathbf{W}}$ be denoted as $\tilde{\mathbf{w}}(k) = (\tilde{\mathbf{w}}_0(k), \tilde{\mathbf{w}}_1(k))$, where $\tilde{\mathbf{w}}_0,\tilde{\mathbf{w}}_1\subset \mathbb{N}^{K}$ and $K \in [N, 2N-1]$. For example, the warp path in Figure~\ref{tkz:shortcommings} is given as [(0,0), (1,0), (2,0), (3,0), (4,1), (5,2), ...]. Essentially, $\tilde{\mathbf{w}}_0$ and $\tilde{\mathbf{w}}_1$ return the indices of $\tilde{\mathbf{f}}$ and $\tilde{\mathbf{g}}$, respectively.

The entries in $\tilde{\mathbf{w}}\in\tilde{\mathbf{W}}$ must meet the following conditions:
\begin{enumerate}
\item Monotonicity and continuity:
\begin{align*}
\tilde{\mathbf{w}}_0(k) \leq &\tilde{\mathbf{w}}_0(k+1) \leq \tilde{\mathbf{w}}_0(k)+1,\\
\tilde{\mathbf{w}}_1(k) \leq &\tilde{\mathbf{w}}_1(k+1) \leq \tilde{\mathbf{w}}_1(k)+1.
\end{align*}
\item Boundary:
\begin{align}\label{eq:boundary0}
\tilde{\mathbf{w}}(0)=(0,0), \tilde{\mathbf{w}}(K-1)=(N-1,N-1).
\end{align}
\end{enumerate}

DTW chooses the warp path that gives the minimum cumulative distance between $\tilde{\mathbf{f}}$ and $\tilde{\mathbf{g}}$. Taking the squared Euclidean Distance as the distance metric, as recommended in \cite{berndt1994}, we get the minimum cumulative distance computed by DTW as  
\begin{align}\label{eq:DTW}
\text{DTW}(\tilde{\mathbf{f}},\tilde{\mathbf{g}})=\min_{\tilde{\mathbf{w}}\in\tilde{\mathbf{W}}} \sum_{k=0}^{K-1}\tilde{\delta}(\tilde{\mathbf{w}}(k)),
%\vspace{-8pt}
\end{align}
where $\tilde{\delta}(\tilde{\mathbf{w_0}}(k),\tilde{\mathbf{w_1}}(k)) =  (\tilde{\mathbf{f}}(\tilde{\mathbf{w}}_0(k))-\tilde{\mathbf{g}}(\tilde{\mathbf{w}}_1(k))^2$.

The computation of $\text{DTW}(\tilde{\mathbf{f}},\tilde{\mathbf{g}})$ is carried out as follows:
\begin{enumerate} 
\item Populate a cumulative distance matrix $\tilde{\mathbf{C}} \subset \mathbb{R}^{N\times N}$. Every point in this matrix gives a value indicating the cheapest path to that point from the start. Every element is given by,

\begin{equation}
\tilde{\mathbf{C}}[i,j] = (\tilde{\mathbf{\delta}}(i,j))+\min 
\begin{cases}
\tilde{\mathbf{C}}[i,j-1],\\
\tilde{\mathbf{C}}[i-1,j-1],\\
\tilde{\mathbf{C}}[i-1,j]\},\\
\end{cases}
\forall i,j \in [0,N-1].
\label{equ:dtw_c}
\end{equation}

\item Backtrack from $\tilde{\mathbf{C}}(N-1,N-1)$ to $\tilde{\mathbf{C}}(0,0)$ to construct the warp path $\tilde{\mathbf{w}}$, where,
\begin{align*}
\tilde{\mathbf{w}}(\tilde{K}-1) = (N-1,N-1), \text{ and}
\vspace{-8pt}
\end{align*}
\begin{multline}
\tilde{\mathbf{C}}[\tilde{\mathbf{w}}(k-1)] { }= \
\min \{\tilde{\mathbf{C}}[\tilde{\mathbf{w}}(k)-\mathbf{d}]\}, \\
\forall \mathbf{d}\in \{(0,1),(1,0),(1,1)\}.
\end{multline}
\end{enumerate} 
The time complexity of DTW is $O(N^2)$, although several algorithms for speeding up the computations exist~\cite{Salvador2007,Silva2016,keogh2005}.

\subsection{Insufficiency for TI applications}\label{sec:shortcomings}
We now move to investigate the characteristics of DTW that prohibit it from being readily used for TI performance evaluation in terms of time and value errors.
These should not be interpreted as limitations of DTW since DTW has not been designed for the question at hand.  
%It is worth remarking that these 
% We recognize three primary limitations of DTW that prevents us from directly employing it is designed for a different purpose, but it provides a method for dealing with time differences and value errors at the same time. This fundamental property makes DTW a suitable starting point. We identify and analyze multiple shortcomings of DTW that make it unsuitable for TI. 
\vspace{2mm}
\subsubsection{Boundary conditions cause unrealistic artifacts}\label{sec:problem_A}
DTW, which measures the similarity between two sequences, is extremely useful for sequences classification problems like correlation power analysis, DNA classification, and notably, speech recognition. In the latter, DTW can be used to correctly recognize a word even when spoken at different speeds or pitches. For achieving this, DTW assumes that every sample in one sequence is related to at least one sample in another sequence. While the boundary conditions in Equation~\eqref{eq:boundary0} ensure that the extreme ends of the sequences are invariably aligned with each other, the monotonicity and continuity conditions ensure that none of the other samples in either sequence is skipped. Figure~\ref{tkz:shortcommings} illustrates the two conditions in action. Segments `I' and `IV' show how the boundary conditions lead to start and end artifacts. On the contrary, for TI applications, this assumption does not hold good as we expect some samples in the ideal signal to be missing or distorted in the observed signal.    
Instead of calculating how two sequences are similar, we want to calculate how the output is different from the input, where differences due to time or value offsets are distinguished.
\input{4_algorithm/length.tex}

\vspace{2mm}
\subsubsection{Unconstrained delay adjustments}\label{sec:problem_B}
Since the objective of DTW is to find the best match between two sequences by minimizing the cumulative the Euclidean distances, as shown in Equation~\eqref{eq:DTW}, delay adjustments are no concern for DTW.
In practice, this means that in pursuit of picking the minimum Euclidean distance, DTW compromises on the delay adjustments.
Indeed, for applications like speech recognition, this is a desirable property, as differences in delay do not matter when identifying words spoken. However, this is unacceptable for TI, as delay and delay adjustments cause more significant degradation in performance than noise in most cases. Examples are presented in Figure~\ref{tkz:shortcommings}. In segment `II', the time derivative of both sequences is low, so that small differences cause large fluctuations in delay. In segment `III', one can see a large number of small changes that lead to minor improvements.

% . There is a penalty for a change in time because every delay adjustment leads to the inclusion of an extra pair of samples. However, both the delay and delay adjustments are not visible at the output. For applications like speech recognition, this is a desirable property, as differences in delay do not matter when identifying words spoken. However, this is unacceptable for TI, as delay and delay adjustments cause more significant degradation in performance than noise in most cases. Examples are presented in Figure~\ref{tkz:shortcommings}. In segment `II', the time derivative of both sequences is low, so that small differences cause large fluctuations in delay. In segment `III', one can see a large number of small changes that lead to minor improvements.
\vspace{2mm}
In the pursuit to identify similar sequences, DTW aims for the highest similarity it can find. It has been proven that DTW achieves its stated goal optimally~\cite{Sakoe1978}. As a consequence, the delay changes are not necessarily recognized at the time instants the real delay changes. In most instances, the changes in delay happen earlier. Multiple examples can be found in Figure~\ref{tkz:shortcommings}. Segments `I' and `IV' start with an adjustment of delay, even though the changes are made toward the end of the corresponding segments. At the start of Segment `II' there is a large delay change a few samples before a small peak that causes the change.

In order to resolve the above issues and design suitable performance measures for TI, we perform substantial refinements to DTW as we describe in the next section.

\section{Design of Quantitative Metric}\label{sec:etvo}
In this section, we begin by presenting the mathematical foundation of \textit{Effective Time- and Value-Offset (ETVO)} framework that sets the stage for proposing the two quantitative metrics for TI. We then move to introducing the metrics -- \textit{Effective Delay-derivative (EDD)} and \textit{Effective RMSE (ERMSE)}.

% We propose four changes to DTW that correspond to the four identified limitations in DTW, as described in Section~\ref{sec:limitations}. The result is a set of measures that can aid the development of TI systems in a significant way.
%
\subsection{Design of ETVO framework} \label{sec:change_A}
We now discuss our refinements for resolving the previously discussed issues of DTW for TI applications. 
% As discussed in Section~\ref{sec:problem_A}, there is a fundamental difference between the purpose of DTW and ETVO. The symmetry in DTW is not needed, and the boundary conditions are undesirable. Thus we try to find a different foundation that is better suited for ETVO.  

% are artifacts at the start and end when applying DTW. These problems stem from the fundamental way DTW operates. DTW tries to match two pieces of signal to each other and give a difference score. This goal is fundamentally different than our goal, where we are interested in the delay of a subset of observed samples. For this goal, there is no requirement for boundaries at the beginning and end of a subset. To properly address this difference a new mathematical base needs to be defined that fits with the task of these measures.
\subsubsection{Relaxation of boundary conditions} \label{sec:change_B}
Let $\mathbf{f}$ and $\mathbf{g}$ denote slices of the input and the output signals, respectively. For ease of explanation, we use the same notations as in DTW, however, without the accent ( $\tilde{}$ ) to denote the ETVO counterparts. We consider a range of possible values for sample delays. The minimum delay is $\Delta T_{min}\in\mathbb{R}$, and the maximum delay is $\Delta T_{max} \equiv \Delta T_{min}+MT$, where $M\subset\mathbb{N}^+$, and $T$ the sample period. Let $N$ denote the length of $\mathbf{g}$. Hence, $\mathbf{f}$ should be of length $N+M-1$. If the first sample of $\mathbf{g}$ is at $t=0$, then the first sample of $\mathbf{f}[k]$ should be at $t=-\Delta T_{min}-(M-1)T$. This is illustrated in Figure~\ref{fig:lengths}.

%
% , we define ETO as the difference in discrete steps of each observed sample to their corresponding ideal sample. From the observed signal, a slice $\mathbf{f}\subset\mathbb{R}^N$ is taken. This slice corresponds to the range of the observed signal for which ETO and EVO will be calculated. The amount of delay values considered is chosen as $M\in\mathbb{N}$. From the ideal signal, a slice $\mathbf{g}\subset\mathbb{R}^{M+N}$ is taken. This slice holds all the values of the ideal signal that $\mathbf{f}$ will be compared to.
% We define an index set $I=\{0,1,2,\cdots,M-1\}$, that contains the indices of the $M$ delay options. We define a warp path $\mathbf{w}\in I^N$ that directly indicates the delay in samples between $\mathbf{f}$ and $\mathbf{g}$ such that the corresponding sample for $\mathbf{f}[n]$ matches the time of $\mathbf{g}[n-\mathbf{w}[n]]$. The slices $\mathbf{f}$ and $\mathbf{g}$ don't have to start at the same point in time, so we define $L$ as the time difference between $\mathbf{f}[0]$ and $\mathbf{g}[0]$, where ETO$ = \mathbf{w}+L$. 
%
Let $\mathbf{W}\subset\mathbb{N}^N$ denote the powerset of possible warp paths to align the output $\mathbf{g}$ onto the input $\mathbf{f}$. The optimal warp path is denoted as $\mathbf{w}\in\mathbf{W}$, where $\mathbf{w}[k]$ indicates that $g[k]$ corresponds to $f[k-w[k]]$. We define ETO as the sample-wise time-offset corresponding to the alignment between $\mathbf{g}$ and $\mathbf{f}$. ETO can be derived directly from the warp path and is expressed as
\begin{equation}
    \text{ETO}[k]=\Delta T_{min}+w[k].
\end{equation}
 We define the Cumulative Distribution Matrix as $\mathbf{C}\subset\mathbb{R}^{N\times M}$, where the $x$-axis indicates the sample index of $\mathbf{g}[k]$, and the $y$-axis is the corresponding ETO. This is illustrated in Figure~\ref{tkz:algorithm}, wherein the value at each entry of $\mathbf{C}$ indicates the cumulative cost of getting to that point. This is similar to the DTW counterpart $\tilde{\mathbf{C}}$ defined in Equation~\eqref{equ:dtw_c} except that the $y$-axis here denotes the ETO. The propagation through $\mathbf{C}$ is 
\begin{equation*}
\mathbf{C}[i,j] = \delta(i,j)+\min 
\begin{cases}
\mathbf{C}[i-1,j],\\
\mathbf{C}[i-1,j-1],\\
\mathbf{C}[i,j+1],
\end{cases}
\end{equation*}
where
\begin{equation*}
    \delta(i,j) \equiv (\mathbf{g}[i]-\mathbf{f}[i-j+M-1])^2, \forall i \in [0,N-1], j \in [0,M-1].
\end{equation*}
\input{4_algorithm/fig_algorithm.tex}
% $\delta(i,j)$ returns the squared Euclidean distance between two samples. 
The three directions for calculating $\mathbf{C}$ correspond directly to the three directions in DTW as defined in Equation~\eqref{equ:dtw_c}. In one step of $\mathbf{g}[k]$, the delay can go down multiple steps, but can only go up one step at a time. For this translated system, the monotonicity and continuity condition is given as
$$0 \leq \mathbf{w}(k+1)\leq \mathbf{w}(k)+1.$$
This translation step accomplishes two things. Firstly, this structure is intuitively appealing for reporting time and value offsets for each output sample. Secondly, unlike in DTW, the search space does not scale quadratically with the signal length. Limiting the search space reduces the memory requirements and accelerates the algorithm. 
With the new foundation in place, we can prevent the start and end samples to have zero delay. The first column of $\mathbf{C}$ is initialized as $\mathbf{C}(0,*)= [0]^M.$ Every starting delay is equally expensive. To remove the ending artifact, we let the last sample of ETO be chosen as the cheapest option, so that
\begin{align*}
\mathbf{C}(N-1,\mathbf{w}[N-1]) &\leq \mathbf{C}(N-1,j), &\forall j\in [0,M-1].
\end{align*}

\subsubsection{Constraining delay adjustments} \label{sec:change_B}
In order to mitigate the issue of unconstrained delay adjustments in DTW, we come up with substantial refinements to its design.
First, let us define what a delay adjustment is in the context of ETVO. It is the change in estimated delay per unit time. If multiple changes of the same type happen consecutively, we treat them as a single change of the cumulative magnitude. Recall that the diagonal arrow and down arrow represent an increase and decrease is delay, respectively. The dark grey squares in Figure~\ref{tkz:algorithm} indicate this.

%Therefore, for every sample, the number of options for delay adjustment increases from three to $M$.
\input{4_algorithm/flowchart.tex}
In the following, we present the mathematical foundation behind cumulative distribution matrix $\mathbf{C}$ and describe the rationale behind the penalties.
\begin{align}
\phantom{C_\uparrow[i,j]}
&\begin{aligned}
\mathllap{C_\rightarrow[i,j]} &= \phantom{C}[i-1,j]
\label{eq:1}
\end{aligned}\\
&\begin{aligned}
\mathllap{C_\downarrow[i,j]} &= \min_{k\subset\mathbb{N}^+}\Big\{\mathbf{C}[i,j+k]\\
&\qquad\quad + \sum_{l=1}^{k-1}\delta(i,j+l) + kP_\text{prop}+P_\text{fixed}\Big\}
\label{eq:2}
\end{aligned}\\
&\begin{aligned}
\mathllap{C\!\text{\tiny{$_\nearrow$}}[i,j]} &= \min_{k\subset\mathbb{N}^+}\Big\{\mathbf{C}[i-k,j-k] + \sum_{l=1}^{k-1}\delta(i-l,j-l)\\
&\qquad\quad + kP_\text{prop}+P_\text{fixed}\Big\}.
\label{eq:3}
\end{aligned}
\end{align}
For every delay adjustment, we introduce two variables -- $P_\text{fixed}$ and $P_\text{prop}$, as shown in Equations~\eqref{eq:2} and \eqref{eq:3}. These correspond to fixed penalty for every delay adjustment and penalty proportional to size of the delay adjustment, respectively. $P_\text{fixed}$ affects the number of delay adjustments, and $P_\text{prop}$ affects the magnitude of each adjustment. Together, these penalties suppress the delay adjustments estimated by the algorithm. The variable $P_\text{prop}$ balances between time and value errors. A high $P_\text{prop}$ reduces the time error and increases the value error, and vice-versa. Hence, ETVO performance approaches DTW as $P_\text{prop}$ tends to zero. As a consequence, $P_\text{prop}$ results in frequent delay adjustments even for minor changes in delay. From the standpoint of TI, this behavior is undesirable, since the human perception is insensitive to smaller delay changes. The addition of $P_\text{fixed}$ controls the frequency of delay adjustments. The best candidate for each of the three directions is calculated as shown in  Equation~\eqref{eq:1}-\eqref{eq:3} and is illustrated in Figure~\ref{tkz:algorithm}.

\subsubsection{Improving the timing of delay adjustments}\label{sec:change_C}
In case of DTW, the delay adjustments do not align with the actual events that trigger the delay changes. For TI, such behavior not only makes analysis hard, but also makes the session quality estimates inaccurate. ETO should not be influenced by an event that occurs in the future. Note that $P_\text{fixed}$ and $P_\text{prop}$ do not address this issue of timing the delay adjustments. Therefore, we propose to introduce slack in delay adjustments where their timing is postponed until the slack penalty $P_\text{slack}$ is breached. $P_\text{slack}$ acts on top of $P_\text{fixed}$ and $P_\text{prop}$ for every delay adjustment, but is only added after an adjustment is made. The addition of $P_\text{slack}$ increases the likelihood that the delay adjustments match the events that cause them. With this, the overall cumulative distribution matrix $\mathbf{C}$ is given as follows.
\begin{multline*}\label{eq:final_C}
\mathbf{C}[i,j] = \delta(i,j)\\
+\begin{cases}
\!C_\rightarrow[i,j] &\text{if } C_\rightarrow[i,j]\!<\!\min\!\Big\{\!C_\downarrow[i,j],C\!\text{\tiny{$_\nearrow$}}[i,j]\!\Big\},\!\!\\
\!C_\downarrow[i,j]\!+\!P_\text{slack} &\text{if } C_\downarrow[i,j]\!<\!\min\!\Big\{\!C_\rightarrow[i,j],C\!\text{\tiny{$_\nearrow$}}[i,j]\!\Big\},\!\!\\
\!C\!\text{\tiny{$_\nearrow$}}[i,j]\!+\!P_\text{slack} &\text{otherwise} \\
\end{cases}
\end{multline*}

\subsubsection{Defining EVO}
Unlike DTW where the residual distance for every sample in the warp path is aggregated into a single number, we represent the value-offset as a time-series that we call EVO. Every sample of EVO indicates the cumulative amount of distance from all samples of $\mathbf{g}$ compared to the corresponding sample in $\mathbf{f}$, excluding the penalties. When ETO increases or stays the same, only one sample of $\mathbf{g}$ is compared to $\mathbf{f}$, however, when ETO reduces, EVO is the cumulative distance between the output sample and several input samples.
This enables obtaining fine-grained information on how samples contribute to the value-offset. The mathematical description for every element of EVO is given by 
\begin{multline*}
    \text{EVO}[i] = 
\begin{cases}
\sum_{l=\text{ETO}[k\!+\!1]}^{\text{ETO}[k]}\delta(i,l) &\text{if } \text{ETO}[i]\!>\!\text{ETO}[i\!+\!1],\\
\delta(i,\text{ETO}[i]) &\text{otherwise.}
\end{cases}
\end{multline*}
Due to this, there are spikes in EVO every time the ETO reduces by a large amount.
\subsubsection{Computational complexity}
\input{4_algorithm/flowchart2.tex}
Besides presenting the ETVO framework, we also provide an efficient way of calculating ETO and EVO. The addition of $P_\text{fixed}$ results in a larger set of values to consider when finding the optimal path. Instead of the three adjacent locations, one has to consider a total of $M$ entries. Besides considering multiple entries, when backtracking to retrieve the delay, one has to consider how many steps were taken. To store that information, we propose  a direction matrix $\mathbf{D}\subset\mathbb{Z}^{M\times N}$. The number stored in $\mathbf{D}(k,i)$ indicates that the next point is at  $i+\mathbf{D}(k,i)$. Because the direction is stored, there is no need to store $\mathbf{C}$ entirely. There are three directions to consider: up, down, and forward. Each direction has one optimal source from which to start. If we remember only the optimal continuations in each stage of the algorithm, we only need to store a value and the corresponding index $C_\downarrow\!\subset\!\mathbb{R}$ and $idx_\downarrow\!\subset\!\mathbb{N}$ for downward propagation, arrays $\!C\!\text{\tiny{$_\nearrow$}}[i,j]\!\subset\!\mathbb{R}^M$ and $idx\!\text{\tiny{$_\nearrow$}}[i,j]\!\subset\!\mathbb{N}^M$ for upward propagation, and array $C_\rightarrow\!\subset\!\mathbb{R}^M$ for forward propagation. The resulting algorithm for populating $\mathbf{D}$ is illustrated with a flow chart in Figure~\ref{tkz:flowchart}. 

The backtracking algorithm needs to account for the multiple steps it can take. The calculation method matches how $\mathbf{C}$ was populated. Figure~\ref{tkz:flowchart2} shows a flow chart of the backtracking algorithm. The complexity of the algorithm for populating $\mathbf{D}$ is determined by two \textit{for loops}. One of the \textit{for loops} is looping through a fixed range $M$, which does not scale with signal length. Therefore the complexity is $\mathcal{O}(N)$. The complexity of the backtracking algorithm is bound by a single \textit{for loop}, so the upper bound on the combined set is also $\mathcal{O}(N)$.

\subsection{Quantitative Metrics for TI}
Based on ETVO, we propose a set of two metrics that can characterize the TI session:
\begin{itemize}
    \item \textit{Effective Delay-Derivative (EDD)} expressed as the average rate of change of ETO, and given as $\frac{1}{N-1}\sum_{k=2}^{N-1}|ETO(k-1)-ETO(k)|$, and
    \item \textit{Effective Root Mean Square Error (ERMSE)} defined as the deviation of value error, and given as $\sqrt{\frac{1}{N}\sum_{k=0}^{N-1} \text{EVO}[k]}$.
\end{itemize} 
The parameters $P_\text{prop}$, $P_\text{fixed}$, and $P_\text{slack}$ influence how EDD weighs against ERMSE. Out of these, $P_\text{prop}$ is the most important of the three. The choice of the parameter values depends on the type of TI application. 

The desirable feature of EDD and ERMSE is the following.
For an application with high dynamic environment, where the movements are reasonably fast, a small amount of noise is insignificant, and hence should be attributed to ERMSE. On the other hand, in low dynamic environment, any amount of noise becomes significant and it makes sense to attribute it to EDD. Therefore, in order to accurately tune $P_\text{prop}$, we have to state how we rate time noise against value noise. One perspective is to consider time noise to be a source of value noise too. The amount of value noise is linearly proportional to the velocity. We suggest using the average velocity of the TI session so that $P_\text{prop} = T\dot{x}_\text{average}$, where $T$ is the sampling period and $\dot{x}_\text{average}$ is the average velocity of the signal.
$P_\text{fixed}$ suppresses unrealistic optimization and micro-adjustments. In practice, we found that a value of $P_\text{fixed}=2P_\text{prop}$ yields good results. $P_\text{slack}$ effectively cleans up the signal and makes the timing more accurate, but it only works if $P_\text{slack}$ is by far the least significant. We found that $P_\text{slack}\leq P_\text{prop}$ yields good results. We believe a more sophisticated method or mathematical backing to tune the penalties is possible in the future.
% For DTW, there is only one source of error that counts: Euclidean distance. For ETVO there are also three penalties to consider: $P_\text{prop}$, $P_\text{fixed}$, and $P_\text{slack}$. These penalties should be weighted properly to balance them against the Euclidean distance. $P_\text{prop}$ should be by far the largest source of penalties during operation. $P_\text{fixed}$ is there to suppress unrealistic optimization and micro-adjustments. In practise we found that a value of $P_\text{fixed}=2P_\text{prop}$ yields good results. $P_\text{slack}$ effectively cleans up the signal and makes the timing more accurate, but it only works if $P_\text{slack}$ is by far the least significant. We found that $P_\text{slack}\leq P_\text{prop}$ yields good results. 

% When using ETVO to diagnose the behavior of a system, we suggest to experiment with a range of values for $P_\text{prop}$ and adjust the other values accordingly, to observe different relationships between delay and noise in the system. As a guideline for choosing $P_\text{prop}$, we recommend Weber's law on sensitivity, where the sensitivity is related to the maximum impulse. From empirical results we found $P_\text{prop}\approx \frac{S}{40}$ to work satisfactory, where $S$ is the maximum impulse. We believe a more sophisticated method to tune the penalties exists, and is future work. \todo[inline]{We should have a good discussion on how to do this}
% \lipsum[1]

%% file: 4_algorithm/length.tex
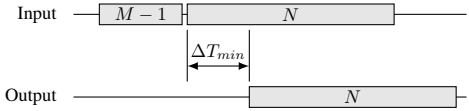
\begin{figure}[t]
\tikzstyle{dot} = [draw,shape=circle,fill=black, scale =.3]
\centering
\begin{tikzpicture}[scale = 0.55, every node/.style={scale=0.7},rotate = 180, xscale = -1]

\def\H{0.25};

\node (input) at (0,0){};
\node (output) at ([shift={(0,2)}] input.center){};
\node [anchor=east] at (input){Input};
\node [anchor=east] at (output){Output};

\draw ([shift={(0.25,0)}] input.center) -- ++ (9.5,0);
\draw ([shift={(0.25,0)}] output.center) -- ++ (9.5,0);

\draw [fill=gray!20] ([shift={(3,-\H)}] input.center) rectangle ++(5,2*\H) node[pos=.5]{$N$};
\draw [fill=gray!20] ([shift={(4.5,-\H)}] output.center) rectangle ++(5,2*\H) node[pos=.5]{$N$};
\draw [fill=gray!20] ([shift={(1.375-0.5,-\H)}] input.center) rectangle ++(2,2*\H) node[pos=.5]{$M-1$};

\draw ([shift={(3,1.5*\H)}] input.center) -- ++ (0,2-3*\H);
\draw ([shift={(4.5,1.5*\H)}] input.center) -- ++ (0,2-3*\H);
\draw [>=latex, <->]([shift={(3,1.25)}] input.center) -- ++ (1.5,0);
\node (L) at ([shift={(3.75,1.25)}] input.center){};
\node [anchor=south] at (L.center){$\Delta T_{min}$};
\end{tikzpicture}
\caption{Illustration of extending the input sequence by $M-1$ samples to avoid the start and end artefacts in ETVO.}
\label{fig:lengths}
\vspace{-8pt}
\end{figure}

%% file: 4_algorithm/fig_algorithm.tex
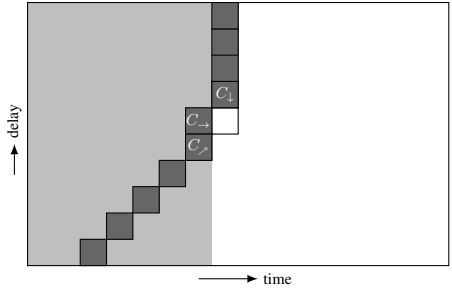
\begin{figure}[t]
\tikzstyle{dot} = [draw,shape=circle,fill=black, scale =.3]
\centering
\begin{tikzpicture}[scale = 0.7, every node/.style={scale=.6},rotate = 180, xscale = -1]

\node (start) at (3.5,2){};
\node (bounds) at (8,5){};

\fill [fill=lightgray] (0,0) rectangle (bounds.center);
\fill [fill=white] (start.center) rectangle ([shift={(0,0)}] bounds.center);
\fill [fill=white] (4,0) rectangle ([shift={(0,0)}] bounds.center);
\draw [fill = white] (start.center) rectangle ([shift = {(.5,.5)}] start.center);
\draw [] (0,0) rectangle (bounds.center);

\foreach \i in {1,...,5}{
% \foreach \i in {1}{
    \fill [black!60] ([shift={(-.5*\i,.5*\i)}] start.center) rectangle ([shift = {(.5-.5*\i,.5+.5*\i)}] start.center);
    \draw [] ([shift={(-.5*\i,.5*\i)}] start.center) rectangle ([shift = {(.5-.5*\i,.5+.5*\i)}] start.center);
}
\draw [fill=black!60] ([shift={(-.5,0)}] start.center) rectangle ([shift = {(0,.5)}] start.center) node[pos=.5, color=white]{$C_\rightarrow$};
\foreach \i in {1,...,4}{
% \foreach \i in {1}{
    \fill [black!60] ([shift={(0,-.5*\i)}] start.center) rectangle ([shift = {(.5,.5-.5*\i)}] start.center);
    \draw [] ([shift={(0,-.5*\i)}] start.center) rectangle ([shift = {(.5,.5-.5*\i)}] start.center);
}
%\draw [fill = darkgray] ([shift={(-2,2.5)}] start.center) rectangle ([shift = {(-1.5,2)}] start.center) node[pos=.5, color=white]{$C\!$\tiny{$_\nearrow$}};
\draw [fill = black!60] ([shift={(-0.5,1)}] start.center) rectangle ([shift = {(-0.0,0.5)}] start.center) node[pos=.5, color=white]{$C\!$\tiny{$_\nearrow$}};

%\draw [fill = darkgray] ([shift={(0,-1)}] start.center) rectangle ([shift = {(.5,-.5)}] start.center) node[pos=.5, color=white]{$C_\downarrow$};
\draw [fill = black!60] ([shift={(0,-.5)}] start.center) rectangle ([shift = {(.5,0)}] start.center) node[pos=.5, color=white]{$C_\downarrow$};

% \draw [>=latex, <-, dashed]([shift={(-1.5,2)}] start.center) -- ([shift={(.25,.25)}] start.center);
% \draw [>=latex, <-, dashed]([shift={(0,.25)}] start.center) -- ([shift={(.25,.25)}] start.center);
% \draw [>=latex, <-, dashed]([shift={(.25,-.5)}] start.center) -- ([shift={(.25,.25)}] start.center);

\node (time) at (4.75,5.25){time};
\draw [>=latex,->] ([shift={(-1.5,0)}] time.center) -- (time);

\node [rotate = 90](delay) at (-.25,2.25){delay};
\draw [>=latex,->] ([shift={(0,1.025)}] delay.center) -- (delay);

\end{tikzpicture}
\caption{The squares indicate all the values that need to be considered to calculate the optimum for the value in the white square. Gray indicates that the optimal value for that spot is already calculated. Dark gray indicates the cheapest source to get to the white square for three different methods. These are the only squares that need to be checked}
\label{tkz:algorithm}
\vspace{-8pt}
\end{figure}

%% file: 4_algorithm/flowchart.tex
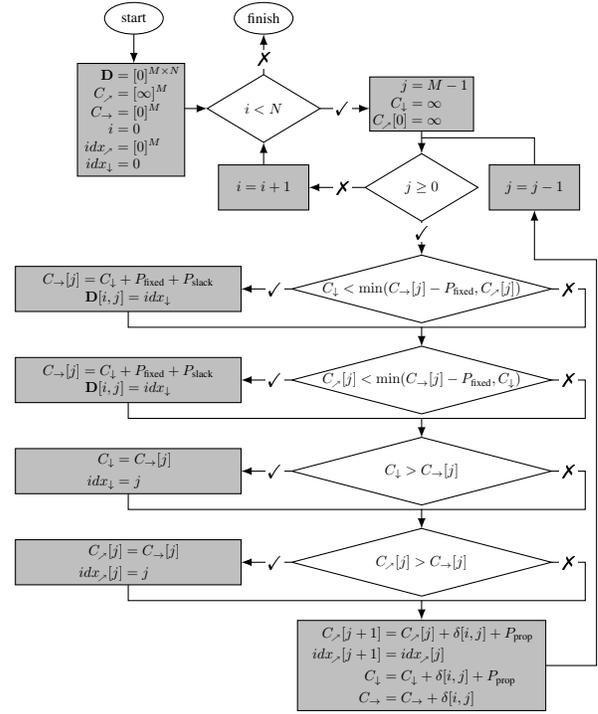
\begin{figure}[t]
\centering
\usetikzlibrary{shapes.geometric}
\begin{tikzpicture}[scale =.60, every node/.style={scale=.55},rotate = 0]

\def\WD{2.875};
\def\HD{.75};

\def\WR{2.5};
\def\HR{0.5};

%------------------------------------------------------------------------------------------------------------------------------------
\node (Q1) at (0,0)                           {};
\node (Q2) at ([shift={(0,-2.02)}] Q1.center) {};
\node (Q3) at ([shift={(0,-2.02)}] Q2.center) {};
\node (Q4) at ([shift={(0,-2.02)}] Q3.center) {};
%------------------------------------------------------------------------------------------------------------------------------------
\foreach \i in {1,2,3,4}
{
\node (A\i) at ([shift={(-6.5,0)}]  Q\i.center) {};

\draw [fill=white] 
   ([shift={(-\WD,0)}]   Q\i.center) 
-- ([shift={(0,\HD)}]    Q\i.center) 
-- ([shift={(\WD,0)}]    Q\i.center)
-- ([shift={(0,-\HD)}]   Q\i.center)
-- ([shift={(-\WD,0)}]   Q\i.center);

\draw [fill=lightgray] ([shift={(-\WR,-\HR)}] A\i.center) rectangle ([shift={(\WR,\HR)}] A\i.center);
\draw [>=latex,->]([shift={(-\WD,0)}] Q\i.center) -- ([shift={(\WR,0)}] A\i.center);
\draw []([shift={(\WD,0)}] Q\i.center) -- ++ (.75,0) |- ([shift={(0,-.85)}] Q\i.center);
\draw [>=latex, ->]([shift={(0,-\HR)}] A\i.center) |-  ([shift={(0,-.85)}] Q\i.center) -- ++ (0,-.43);

\node (checkmark) at ([shift={(-3.25,0)}] Q\i.center) {};
\fill [fill=white] ([shift={(-.2,-.2)}] checkmark.center) rectangle ([shift={(.2,.2)}] checkmark.center);
\node at (checkmark.center) {\Checkmark};

\node (fail) at ([shift={(3.25,0)}] Q\i.center) {};
\fill [fill=white] ([shift={(-.2,-.2)}] fail.center) rectangle ([shift={(.2,.2)}] fail.center);
\node at (fail.center) {\XSolidBrush};
}
%------------------------------------------------------------------------------------------------------------------------------------
\node at (Q1.center) {$C_\downarrow < \min(C_\rightarrow[j]-P_\text{fixed},C\!\text{\tiny{$_\nearrow$}}[j])$};
\node at (Q2.center) {$C\!\text{\tiny{$_\nearrow$}}[j] < \min(C_\rightarrow[j]-P_\text{fixed},C_\downarrow )$};
\node at (Q3.center) {$C_\downarrow > C_\rightarrow[j]$};
\node at (Q4.center) {$C\!\text{\tiny{$_\nearrow$}}[j] > C_\rightarrow[j]$};

\node [align = center] at (A1.center) {$C_\rightarrow[j] = C_\downarrow+P_\text{fixed}+P_\text{slack}$\\$\mathbf{D}[i,j] = idx_\downarrow$};
\node [align = center] at (A2.center) {$C_\rightarrow[j] = C_\downarrow+P_\text{fixed}+P_\text{slack}$\\$\mathbf{D}[i,j] = idx_\downarrow$};
\node [align = center] at (A3.center) {$\begin{aligned}C_\downarrow &= C_\rightarrow[j]\\idx_\downarrow&=j\end{aligned}$};
\node [align = center] at (A4.center) {$\begin{aligned}C\!\text{\tiny{$_\nearrow$}}[j] &= C_\rightarrow[j]\\idx\!\text{\tiny{$_\nearrow$}}[j]&=j\end{aligned}$};
%------------------------------------------------------------------------------------------------------------------------------------
\node [] (S3) at ([shift={(0,-8.35)}] Q1.center) {};
\draw [fill=lightgray] ([shift={(-2.75,-1)}] S3.center) rectangle ([shift={(2.75,1)}] S3.center);
\node [align = center] at (S3.center) {$
\begin{aligned}
%\delta &= \|f[k]-g[k-i+M]\|\\
C\!\text{\tiny{$_\nearrow$}}[j+1]&=C\!\text{\tiny{$_\nearrow$}}[j]+\delta[i,j]+P_\text{prop}\\
idx\!\text{\tiny{$_\nearrow$}}[j+1] &= idx\!\text{\tiny{$_\nearrow$}}[j]\\
C_\downarrow&=C_\downarrow+\delta[i,j]+P_\text{prop}\\
C_\rightarrow&=C_\rightarrow+\delta[i,j]\\
\end{aligned}$};
%------------------------------------------------------------------------------------------------------------------------------------
\node (F2) at ([shift={(0,2.25)}] Q1.center){};
\draw [fill=white] 
   ([shift={(-1.25,0)}]   F2.center) 
-- ([shift={(0,\HD)}]    F2.center) 
-- ([shift={(1.25,0)}]    F2.center)
-- ([shift={(0,-\HD)}]   F2.center)
-- ([shift={(-1.25,0)}]   F2.center);
\node () at (F2.center){$j\geq0$};

\node (F1) at ([shift={(-3.5,1.75)}] F2.center){};
\draw [fill=white] 
   ([shift={(-1.25,0)}]   F1.center) 
-- ([shift={(0,\HD)}]    F1.center) 
-- ([shift={(1.25,0)}]    F1.center)
-- ([shift={(0,-\HD)}]   F1.center)
-- ([shift={(-1.25,0)}]   F1.center);
\node () at (F1.center){$i<N$};

\node (S1) at ([shift={(0,0.1)}] F1.center -| F2.center){};
\node (S2) at (F1.center |- F2.center){};

\draw [fill=lightgray] ([shift={(-1.15,-\HR-0.1)}] S1.center) rectangle ([shift={(1.15,\HR+0.1)}] S1.center);
\node [align=center] at (S1.center){
$\begin{aligned}
j&=M-1\\[-3pt]
C_\downarrow&=\infty\\[-3pt]
C\!\text{\tiny{$_\nearrow$}}[0]&=\infty
\end{aligned}$
};

\draw [fill=lightgray] ([shift={(-1,-\HR)}] S2.center) rectangle ([shift={(1,\HR)}] S2.center);
\node at (S2.center){$i=i+1$};
%------------------------------------------------------------------------------------------------------------------------------------
\node (S4) at ([shift={(2.5,0)}] F2.center){};
\draw [fill=lightgray] ([shift={(-1,-\HR)}] S4.center) rectangle ([shift={(1,\HR)}] S4.center);
\node at (S4.center){$j=j-1$};
\draw [>=latex,->] ([shift={(2.75,0)}] S3.center) -| ++ (1.125,9)-| ([shift={(0,-\HR)}] S4.center);
\draw [] ([shift={(0,\HR)}] S4.center) |- ([shift={(0,1.1)}] F2.center);
%------------------------------------------------------------------------------------------------------------------------------------
\draw [>=latex,->] ([shift={(0,-\HD)}] F2.center) -- ([shift={(0,\HD)}] Q1.center);
\draw [>=latex,->] ([shift={(-1.25,0)}] F2.center) -- ([shift={(1,0)}] S2.center);
\draw [>=latex,<-] ([shift={(0,\HD)}] F2.center) -- ([shift={(0,-\HR-0.1)}] S1.center);

\node (checkmark) at ([shift={(0,-1)}] F2.center) {};
\fill [fill=white] ([shift={(-.2,-.2)}] checkmark.center) rectangle ([shift={(.2,.2)}] checkmark.center);
\node at (checkmark.center) {\Checkmark};

\node (fail) at ([shift={(-1.75,0)}] F2.center) {};
\fill [fill=white] ([shift={(-.2,-.2)}] fail.center) rectangle ([shift={(.2,.2)}] fail.center);
\node at (fail.center) {\XSolidBrush};
%------------------------------------------------------------------------------------------------------------------------------------
\draw [>=latex,->] ([shift={(1.25,0)}] F1.center) -- ([shift={(-1.15,-0.1)}] S1.center);
\draw [>=latex,->] ([shift={(0,\HD)}] F1.center) -- ([shift={(0,1.65)}] F1.center);
\draw [>=latex,<-] ([shift={(0,-\HD)}] F1.center) -- ([shift={(0,\HR)}] S2.center);

\node (checkmark) at ([shift={(1.75,0)}] F1.center) {};
\fill [fill=white] ([shift={(-.2,-.2)}] checkmark.center) rectangle ([shift={(.2,.2)}] checkmark.center);
\node at (checkmark.center) {\Checkmark};

\node (fail) at ([shift={(0,1.1)}] F1.center) {};
\fill [fill=white] ([shift={(-.2,-.2)}] fail.center) rectangle ([shift={(.2,.2)}] fail.center);
\node at (fail.center) {\XSolidBrush};
%------------------------------------------------------------------------------------------------------------------------------------
\node (finish) at ([shift={(0,2)}] F1.center){finish};
\draw (finish.center) ellipse (.65 and .35);

\node (start) at ([shift={(-2.875,2)}] F1.center){start};
\draw (start.center) ellipse (.65 and .35);

\node (S5) at ([shift={(-2.875,-.25)}] F1.center){};
\draw [fill=lightgray] ([shift={(-1.25,-1.25)}] S5.center) rectangle ([shift={(1.125,1.25)}] S5.center);
\node [align=center] at (S5.center){
$\begin{aligned}
\mathbf{D}&=[0]^{M\times N}\\[-3pt]
C\!\text{\tiny{$_\nearrow$}}&=[\infty]^M\\[-3pt]
C_\rightarrow&=[0]^M\\[-3pt]
i&=0\\[-3pt]
idx\!\text{\tiny{$_\nearrow$}}&=[0]^M\\[-3pt]
idx_\downarrow&=0
\end{aligned}$
};

\draw [>=latex, ->] ([shift={(0,-.35)}] start.center) -- ([shift={(0,1.25)}] S5.center);
\draw [>=latex, <-] ([shift={(-1.25,0)}] F1.center) -- ++ (-.5,0);

%------------------------------------------------------------------------------------------------------------------------------------

\end{tikzpicture}
\caption{Flowchart of the algorithm that finds the optimal way of traversing the delay, given the constraints specified for ETO.} 
\label{tkz:flowchart}
\vspace{-8pt}
\end{figure}

%% file: 4_algorithm/flowchart2.tex
\begin{figure}[]
\centering
\usetikzlibrary{shapes.geometric}
\begin{tikzpicture}[scale =.65, every node/.style={scale=.65},rotate = 0, xscale=-1]

\def\WD{1.25};
\def\HD{.75};

\def\WR{2.5};
\def\HR{0.5};

\node (Q1) at (0,0){};

%------------------------------------------------------------------------------------------------------------------------------------

\node (F0) at ([shift={(0,2.25)}] Q1.center){$i>=0$};
\node (F1) at ([shift={(0,0)}] Q1.center){$l>0$};
\node (F2) at ([shift={(0,-2.25)}] Q1.center){$D[i,j]<0$};
\node (F3) at ([shift={(0,-4.5)}] Q1.center){$D[i,j]>0$};

\foreach \i in {1,2,3}{
\draw [] 
   ([shift={(-\WD,0)}]   F\i.center) 
-- ([shift={(0,\HD)}]    F\i.center) 
-- ([shift={(\WD,0)}]    F\i.center)
-- ([shift={(0,-\HD)}]   F\i.center)
-- ([shift={(-\WD,0)}]   F\i.center);
\draw [>=latex,->] ([shift={(\WD,0)}] F\i.center) -- ++ (1,0);
\node (checkmark) at ([shift={(1.6,0)}] F\i.center) {};
\fill [fill=white] ([shift={(-.2,-.2)}] checkmark.center) rectangle ([shift={(.2,.2)}] checkmark.center);
\node at (checkmark.center) {\Checkmark};

\draw [>=latex,->] ([shift={(0,-\HD)}] F\i.center) -- ([shift={(0,-2.25+\HD)}] F\i.center);
\node (fail) at ([shift={(0,-1.01)}] F\i.center) {};
\fill [fill=white] ([shift={(-.2,-.2)}] fail.center) rectangle ([shift={(.2,.2)}] fail.center);
\node at (fail.center) {\XSolidBrush};

\draw [fill=lightgray] ([shift={(\WD+1,\HR)}] F\i.center) rectangle ++ (\WR,-2*\HR);
\draw ([shift={(4.75,0)}] F\i.center) -- ([shift={(5.5,0)}] F\i.center);
}

\node [align=center] at ([shift={(3.5,0)}] F1.center){$l=l-1$\\$j=j-1$};
\node [align=center] at ([shift={(3.5,0)}] F2.center){$l=-\mathbf{D}[i,j]$};
\node [align=center] at ([shift={(3.5,0)}] F3.center){$j=j+\mathbf{D}[i,j]$};

\node (S1) at (0,-6.5){};
\draw [fill=lightgray] ([shift={(-\WR*2/3,-\HR)}] S1.center) rectangle ([shift={(\WR*2/3,\HR)}] S1.center);
\node [align=center] at (S1.center) {$\text{ETO}[i]=j-\Delta T_{min}$\\$i=i-1$};
\draw [>=latex, <-] ([shift={(\WR*2/3,0)}] S1.center) -| ++ (4.25-\WR/6, 6.5);
\draw [>=latex, ->] ([shift={(-\WR*2/3,0)}] S1.center) -| ++ (-1,6.5) |- ([shift={(-\WR/2,0)}] F0.center);

%\draw [>=latex, <-] ([shift={(\WR/2,0)}] S1.center) -| ++ (4.25,6.35);

\draw [] 
   ([shift={(-\WD,0)}]   F0.center) 
-- ([shift={(0,\HD)}]    F0.center) 
-- ([shift={(\WD,0)}]    F0.center)
-- ([shift={(0,-\HD)}]   F0.center)
-- ([shift={(-\WD,0)}]   F0.center);
\draw [>=latex,->] ([shift={(\WD,0)}] F0.center) -- ++ (1,0);
\node (checkmark) at ([shift={(1.6,0)}] F0.center) {};
\fill [fill=white] ([shift={(-.2,-.2)}] checkmark.center) rectangle ([shift={(.2,.2)}] checkmark.center);
\node at (checkmark.center) {\XSolidBrush};
\draw [>=latex,->] ([shift={(0,-\HD)}] F0.center) -- ([shift={(0,-2.25+\HD)}] F0.center);
\node (fail) at ([shift={(0,-1.01)}] F0.center) {};
\fill [fill=white] ([shift={(-.2,-.2)}] fail.center) rectangle ([shift={(.2,.2)}] fail.center);
\node at (fail.center) {\Checkmark};

\node (S2) at (0,4.35){};
\draw [fill=lightgray] ([shift={(-\WR,-1.75*\HR)}] S2.center) rectangle ([shift={(\WR,1.75*\HR)}] S2.center);
\node [align=center] at (S2.center) {$i=N-1$\\$\text{ETO}=[0]^N$\\$l=0$\\$j=j_{start}$};
\draw [>=latex, <-] ([shift={(0,\HD)}] F0.center) -- ++ (0,.5);

\node (finish) at ([shift={(2.9,0)}] F0.center){finish};
\draw (finish.center) ellipse (.65 and .35);

\node (start) at ([shift={(3.75,2.1)}] F0.center){start};
\draw (start.center) ellipse (.65 and .35);
\draw [>=latex, ->] ([shift={(-.65,0)}] start.center) -- ++ (-.6,0);
\end{tikzpicture}
\caption{Flowchart of the backtracking algorithm used to extract the ETO from direction matrix $\mathbf{D}$}
\label{tkz:flowchart2}
\vspace{-8pt}
\end{figure}
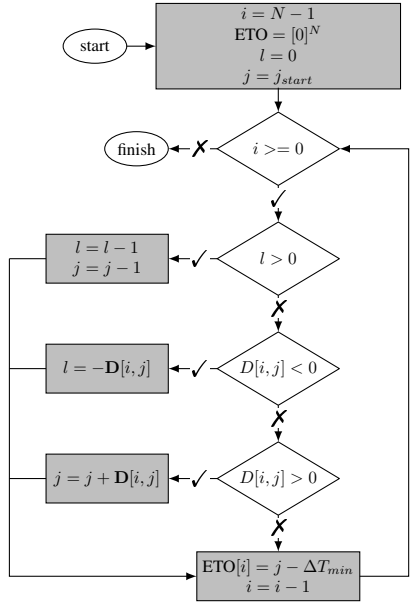

%% file: 5_experimentalsetup/experimentalsetup.tex
\section{Experimental setup}\label{sec:setup}
To validate our proposed metric, we develop a realistic TI testbed, where a human user can interact with a remotely rendered virtual environment (VE) over a network. We consider a recently proposed testbed for simulating TI interaction (Section~\ref{subsec:stdsetup}). This testbed lacks a network and is therefore not suited for mimicking TI interactions. Hence we carry out significant refinements to realize a networked TI testbed that is fast enough to provide a comfortable TI experience (see Section~\ref{subsec:newsetup}).

\subsection{Standard TI testbed} \label{subsec:stdsetup}
The standard TI testbed proposed recently by Bhardwaj et al., simulates a TI session by having the human user interact with a Virtual Environment (VE) in slave domain via haptic and visual feedback  \cite{Bhardwaj2017}. The master domain transmits position information of the haptic device sampling rate of \SI{1}{kHz} to the slave domain. 
The resulting calculated force is sent back to the master domain along with the visual rendering of VE. A schematic overview of this setup is indicated as the gray blocks in Figure~\ref{fig:setup}. In this figure, both master and slave domains are collocated for ease of experimentation. The haptic device used in this setup is a Novint Falcon \cite{novintfalcon}. Force calculation and rendering in the VE is implemented using Chai3D \cite{Chai3dfeatures}.

\subsection{Networked TI testbed}\label{subsec:newsetup}
\begin{figure}[t]
\centering
\includegraphics[width=0.499\textwidth]{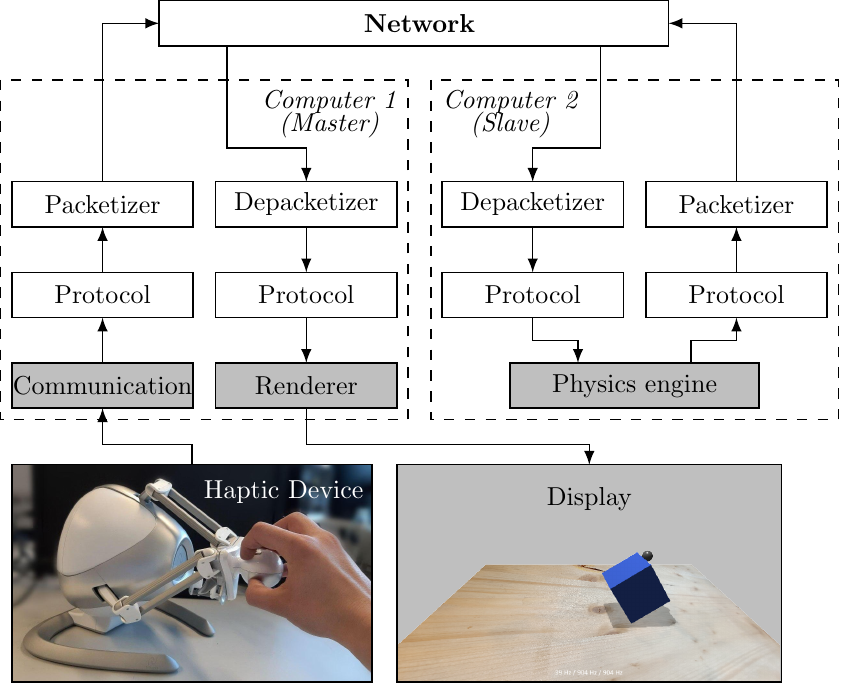}
\caption{Schematic overview of our measurement setup. The components present in the reference setup of Bhardwaj et al. highlighted in gray \cite{Bhardwaj2017}. }
\label{fig:setup}
\vspace{-8pt}
\end{figure}

We extend the testbed by decoupling the master and slave domains (each consisting of a workstation) across a network. Therefore, (i) the VE needs to be rendered on a separate workstation, and (ii) the two workstations are connected via a physical network. 
The local simulation is split into a renderer at the master domain and a physics engine at the slave domain. The physics engine translates force into movement and sends that as feedback.
The real-network conditions between the master and slave are emulated using Netem on a separate workstation running Ubuntu 18.04~\cite{netem}. Netem can emulate network parameters such as the latency, jitter, and packet losses.
A schematic overview of the entire system is shown in Figure \ref{fig:setup}. 
To emulate real network conditions, we use the Gilbert-Elliott model to handle packet loss, which is known to mimic real network behavior closely \cite{Hasslinger2008}. For delay jitter, we use a high correlation factor of \SI{90}{\%} between latency of the packets. We specify the configured parameters of the loss model, average latency, jitter, and the algorithm parameters $P_\text{prop}$, $P_\text{fixed}$, and $P_\text{slack}$ in Section~\ref{sec:results} as and when required. 

%% file: 6_performanceanalysis/performanceanalysis.tex
% \clearpage
%
%
% Scenario 1: deadband (0.15cm) and gemodel (5% loss, with 50% burst chance)
% Scenario 2: 15 ms ping, 10 ms jitter no loss
% Scenario 3: 15 ms ping, 10 ms jitter no loss
% Scenario 4: 15 ms ping, 1 ms jitter no loss

\section{Performance analysis}\label{sec:results}
\begin{figure*}[t]
    \centering
    \includegraphics[width=\textwidth]{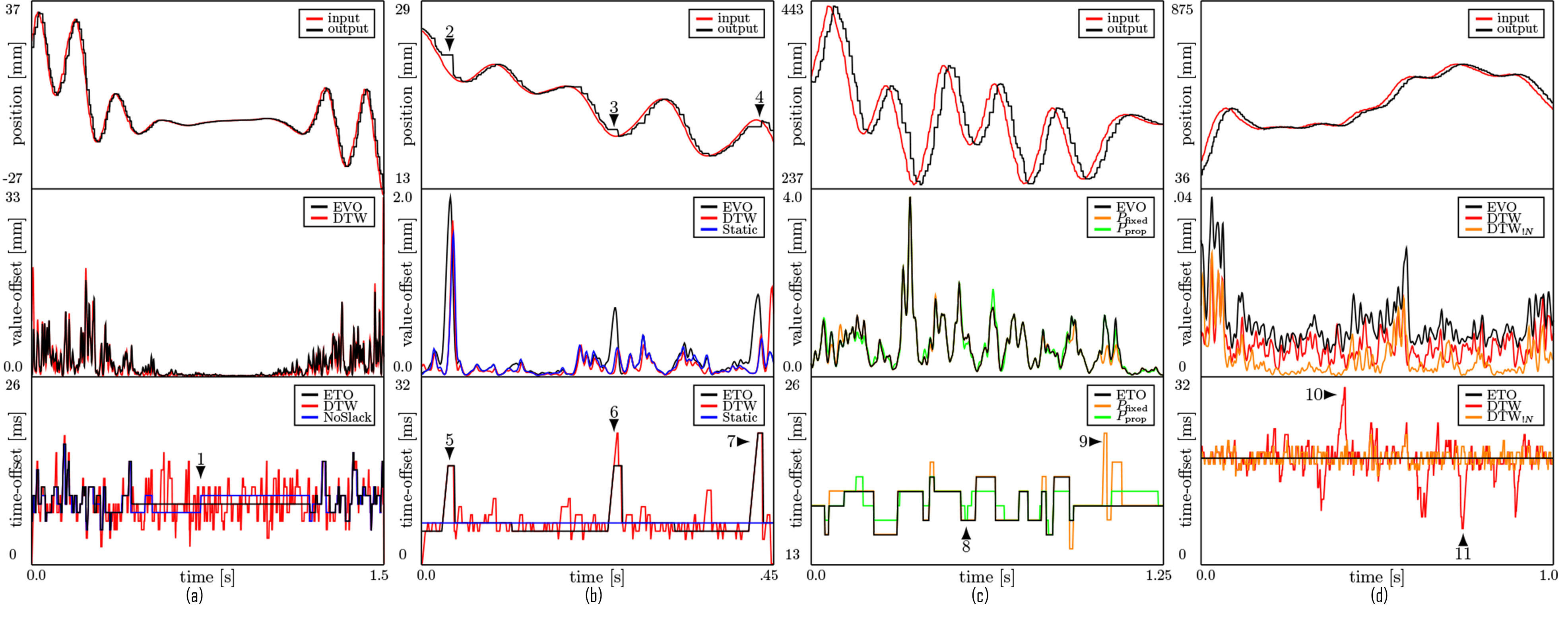}
    \caption{Comparison of the performances of DTW and ETVO frameworks in a variety of experimental setups under the conditions (a) Heavily varying position signal, (b) Packet losses and perceptual deadband, (c) Effects of $P_\text{prop}$ and $P_\text{fixed}$, and (d) Effect of addition of noise to input signal.}
    \label{fig:results}
    \vspace{-8pt}
\end{figure*}
We conduct rigorous experimentation to comprehensively evaluate our metrics using the networked TI testbed. For the sake of simplicity, we consider only one axis of the position signal corresponding to the haptic device. Henceforth, we will refer to that axis signal as input. We compare the performances of DTW and ETVO. We formulate several scenarios which enable easy understanding of the working of ETVO. Finally, we demonstrate the strength of the proposed performance metrics along with describing how they can be arrived at based on the ETVO.
%
% \vspace{-8pt}
% \subsection{Contradicting inferences between QoE and QoS metrics}
% We choose HPW-PSNR \cite{Sakr2007} as the QoE metric, and latency as the QoS metric. Note that similar scenarios can be observed for other combinations of QoE and QoS metrics. Using our testbed, We record the position trace of the haptic device during the TI interaction. 
% In order to visualize how two different TI systems could alter the input, we add a latency of \SI{15}{ms} and \SI{50}{ms} to the input signal.
% %as shown in Figure.~\ref{fig:delayedsignals}. 
% For all practical purposes, this is equivalent to the latency induced by the TI systems. Note that for the sake of brevity, we are only showing a small fragment of the complete signal.

% As per the latency, it is quite obvious that the former TI system outperforms the latter since it introduces lower latency.
% Now, we measure the HPW-PSNR of the two output signals with respect to the input signal.
% We observe that the both systems have an equal HPW-PSNR of \SI{9.21}{}. This is due to the fact that adding a static latency to a signal makes no difference to the PSNR. Going by HPW-PSNR, the performances of the two systems are identical which completely contradicts the inference drawn on the basis of latency. Hence, existing metrics can often lead to ambiguous situations, thereby reinforcing the need for the proposed quantitative metrics that can better represent the TI system performance. 
% \todo[inline, author=Kees]{do we want to make this more convincing? We are already doing something interesting with the measure plot}
%
\subsection{Comparison between DTW and ETVO}
For this work we made alterations to the base DTW algorithm. In this section, we take a look at the effects of these alterations. To illustrate the differences in behavior between DTW and ETVO, we picked four fragments of time, each corresponding to a different signal. The fragments are picked to illustrate different aspects of decision making between DTW and ETVO.
We start by gauging the sensitivity of each of the schemes to the signal variations. For this experiment, we set the following configuration: 
\textbf{[average lat., jitter, $P_\text{prop}$, $P_\text{fixed}$] = [\SI{15}{ms}, \SI{10}{ms}, \SI{0.01}{}, \SI{0.005}]}.
Figure~\ref{fig:results}(a) corresponds to this experimental setup. It can be observed that when the velocity is high (extremes of the plot), DTW shows vigorous fluctuations in time-offset estimation. On the other hand, ETVO sensibly changes its estimation of ETO exhibiting resilience to the noise in the signal. When the velocity is low in the middle, ETVO being context-aware stops adjusting the ETO since there is a negligible improvement in EVO. DTW however, does not care about how minor changes and keeps constantly adjusting the time-offset, irrespective of the context. The scenario for $P_\text{slack}=0$ is also demonstrated (labelled as 'NoSlack'). In this signal, a change in delay is randomly made in the middle of the quiet period (as indicated with 1), while the ETO with $P_\text{slack}$ postpones that decision to a more sensible moment. Note that in the value domain, ETVO and DTW perform similar, despite the significantly higher number of delay adjustments performed by DTW. Moreover, DTW lets the delay fluctuate regardless of what is going on, while ETVO modulates the delay only at high velocities. ETVO reasonably concludes that a higher update rate is desired when more is happening. This example shows how ETVO makes evaluations that are context-aware.  

We now investigate the ETVO performance in the presence of fewer haptic updates. This could be a consequence of either packet losses or employment of compression schemes. To capture the effects of these phenomena, we emulate this scenario through two processes: perceptual deadband \cite{Vittorias2009} and packet losses. This enables us to evaluate the performance when the value-offset is considerable. We choose a deadband of \SI{5}{\%} and bursty loss following the Gilbert-Elliott model with parameters $p=5\%$ and $r=50\%$, as prescribed in \cite{Ellis2014}. For this experiment, we set the following configuration:
\textbf{[average lat., jitter, $P_\text{prop}$, $P_\text{fixed}$] = [\SI{00}{ms}, \SI{00}{ms}, \SI{0.005}{}, \SI{0.005}]}.
Figure~\ref{fig:results}(b) corresponds to this scenario.
It can be clearly seen that the output signal is subject to a noticeable jitter due to the combined effect of packet loss and deadband.
However, there are three specific instances (indicated by markers 2-4) where the combined effect of deadband and bursty losses prominently results in no change in the output signal.
In this case, DTW relentlessly adjusts the time-offset as the deadband and losses are slightly degrading the signal. 
ETVO, however, remains robust to the jitter for the most part and sits at the average delay. But in cases where the effect is prominent, ETO is adjusted as indicated with markers 5-7. 
The value-offset is smoothed with a Gaussian distribution for visual clarity. 
As one expected, in the value domain ETVO is slightly higher than DTW. Although each of the schemes perform their corresponding tasks accurately, the behavior that is most favorable for TI applications is exhibited by ETVO. Especially considering that time variations are considered to be extremely detrimental to the session quality.

In this part we show the distinct effects of $P_\text{prop}$ or $P_\text{fixed}$, and the need of considering both. We show the results in Figure~\ref{fig:results}(c). 
We consider three different settings for algorithm parameters: \\(i) \textbf{[$P_\text{prop}$, $P_\text{fixed}$]} = [0.025,0.05] (corresponds to black curve), \\(ii) \textbf{[$P_\text{prop}$, $P_\text{fixed}$]} = [0.05,0] (corresponds to amber curve), and \\(iii) \textbf{[$P_\text{prop}$, $P_\text{fixed}$]} = [0,0.1] (corresponds to green curve).  Marker 7 indicates an event where the ETO with $P_\text{fixed}=0$ adjusts in multiple small steps. In this case, there is no difference in extra cost associated with using multiple steps, causing the amount of adjustments to increase. Marker 8 indicates an event where the ETO with $P_\text{prop}=0$ causes a large step changes but limited in the number of steps. In this case, there is no extra cost associated with how much the ETO changes, giving rise to giving rise to unreasonably large ETO adjustments. The signal that has both components has a similar performance in the value domain, but a significantly less cluttered ETO. 

Figure~\ref{fig:results}(d) shows the effect that high-frequency noise has on DTW and ETVO. For this purpose, we introduce AWGN with an SNR of \SI{70}{dB}. The experimental configuration is as follows:
\textbf{[average lat., jitter, $P_\text{prop}$, $P_\text{fixed}$, $P_\text{slack}$] = [\SI{15}{ms}, \SI{1}{ms}, \SI{0.005}{}, \SI{0.01}{}, \SI{0.005}{}]}.
Both DTW and EVO are plotted with the noise added, while $\text{DTW}_{!N}$, is a version of DTW without the added AWGN. High-frequency noise is a good example of a common way of signal distortion that DTW cannot deal with properly. Note that ETO outperforms the best case DTW i.e. $\text{DTW}_{!N}$, demonstrating its noise resilience. Further, one can also notice the vulnerability of DTW to even a marginal amount of noise causing time-offset to fluctuate vigorously.  
\subsection{Performance of EDD and ERMSE}
\begin{figure}[t]
    \centering
    \includegraphics[width=0.499\textwidth]{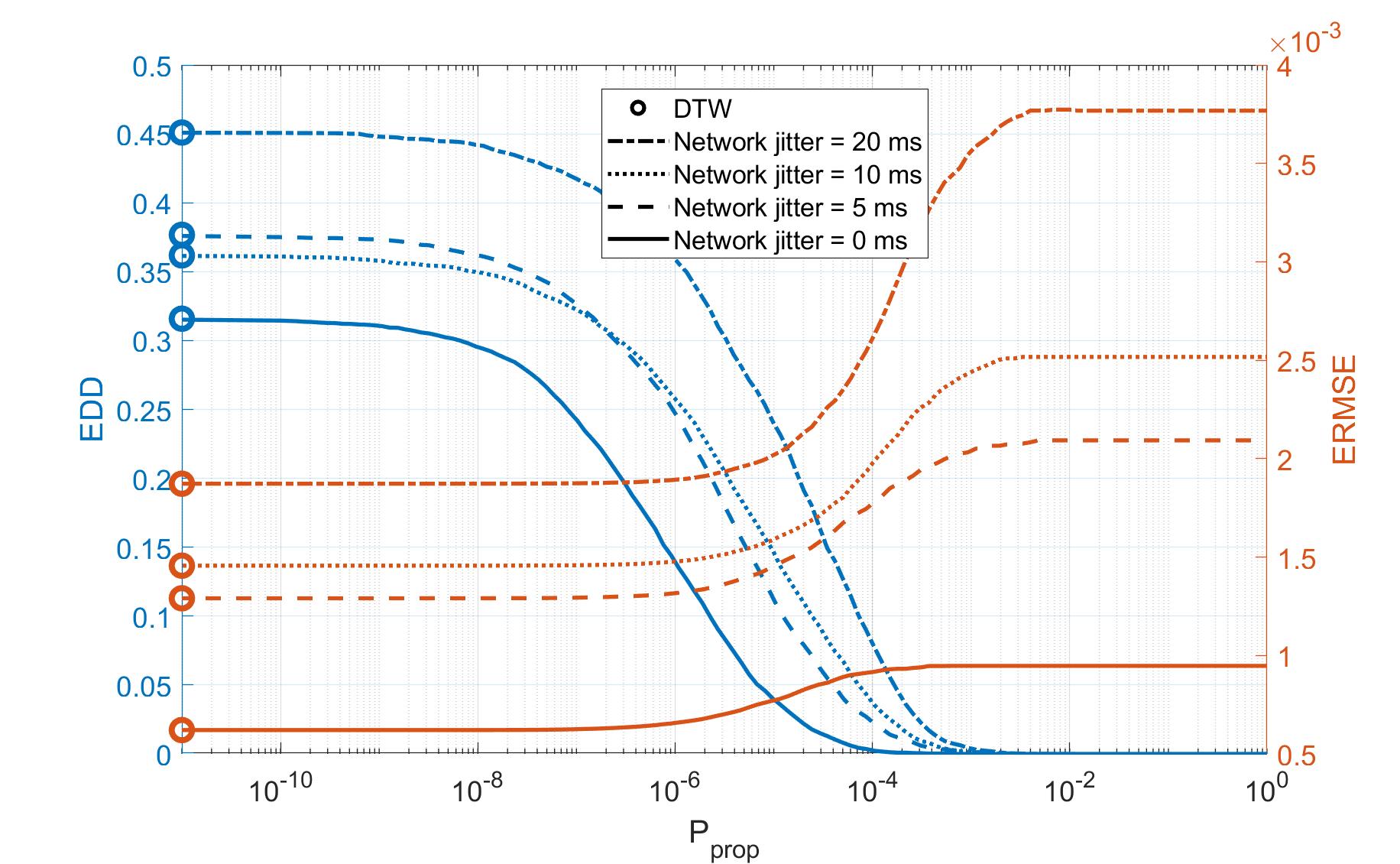}
    \caption{Based on the data of our experiments, the EDD and ERMSE that ETVO calculates are plotted as a function of $P_\text{prop}$.} %The markers in the center indicate the tuned $P_\text{prop}$ for each of the experiments based on average velocity, and the resulting EDD an ERMSE at those point. These markers come in pairs that belong to the same data set.}
    % \caption{This figure plots the reported delay-derivative and RMSE for different experiments that are conducted using the Novint Falcon and Netem to simulate varying amounts of network jitter. The DTW point is generated using ETVO with $P_\text{prop} = 0$. This behaves as DTW without boundary conditions. Including the boundary conditions will add a varying amount of delay-derivative and RMSE. The RMSE point is generated using ETVO with $P_\text{prop} = \infty$. This creates a constant time-offset the leads to the lowest RMSE with zero delay-derivative. However, Existing measures cannot find this time-offset, so they perform significantly worse, depending on the time-offset and velocity in the signal.}
    \label{fig:pprop_sweep}
\end{figure}
% \todo[inline]{I was thinking about this for quite a while. The point is, we can just explain what DTW and RMSE are going to do. They will literally pick where those curves end up on the edges... or worse. Now what is left is writing the story}
In previous subsections, we demonstrated how the performance of ETVO is desirable for TI in comparison to DTW. In this section, we demonstrate how the proposed measures EDD and ERMSE perform. We conduct four runs of the experiment with one value for following network jitter in each run -- \SI{0}{ms}, \SI{5}{ms}, \SI{10}{ms}, and \SI{20}{ms}. In all experiments, the input and output signal durations are larger than \SI{10}{s}. In each run, we calculate the EDD and ERMSE for a wide range of $P_\text{prop}$ to demonstrate how the penalties balance out the EDD and ERMSE. The results are shown in Figure~\ref{fig:pprop_sweep}. 

DTW assigns no value to EDD, which means $P_\text{prop}=0$. The endpoints on the left match what DTW would report, but without accounting for the starting and the ending artifacts. When $N\to\infty$ the contribution of those artifacts become negligible, but for small $N$, the contributions can be significant in comparison with ERMSE. For high $P_\text{prop}$, the EDD becomes zero, as any change is too expensive. However, we can still have any constant delay, so the result is the ERMSE value for the optimal constant delay. This can be seen as the best scenario for ERMSE when the optimal constant delay is known. Existing measures do not compute this, so in practise the ERMSE will be orders of magnitude higher for the methods proposed in literature.
For every experiment there is a pair of lines that indicates the balance between EDD and ERMSE for $P_\text{prop}\in[\approx 0,1]$. For low values of $P_\text{prop}$, the signals reach a point where decreasing $P_\text{prop}$ no longer leads to a reduction in ERMSE and therefore an increase in EDD. One can also see the \textit{law of diminishing returns} in play, where a large increase in amount  EDD is required for the same amount of ERMSE reduction. For high values of $P_\text{prop}$, signals reach a point where increasing $P_\text{prop}$ no longer leads to a decrease in EDD. That is because at that point EDD is already zero. Again, we see the law of diminishing returns, where the ERMSE goes up quite a bit for only a marginal reduction in EDD. 
Both DTW and taking RMSE of the output directly (constant delay)   attempt to account all errors leading a single metric such as mean error. As we see in Figure~\ref{fig:pprop_sweep}, DTW takes one extreme and constant delay takes another extreme. However, we believe that the optimal should be in between these extremes. ETVO provides a handle in the form of tunable penalties ($P_\text{prop}$) using which one can select the tolerable delay and corresponding least error. 

%% file: 7_conclusion/conclusion.tex
\section{Conclusion and future work}\label{sec:conclusions}
As the field of Tactile Internet (TI) is advancing fast, there is a strong need for quantifying performance sessions objectively. In this paper, we first presented first lack of TI performance metrics in the state of the art and their limitations for using them to characterize TI sessions and systems. Specifically both time offsets and value offsets in the input signal and the received (or output) signal cannot be extracted separately in those methods. To overcome these limitations, we find Dynamic Time Warping (DTW) algorithm used in speech recognition as the most suitable. However, we also encountered certain issues in applying DTW directly to characterize performance of TI applications. We demonstrated the issues and  provided a  mathematical framework to address them. During the course, we proposed two novel concepts: Effective Time-Offset (ETO) and Effective Value-Offset (EVO) that enable the representation of TI performance at a fine scale. Based on ETO and EVO, we defined two quantitative performance metrics - Effective Delay-Derivative (EDD) and Effective RMSE (ERMSE) - for quantifying the impact of network effects on the quality of a TI session. 

Through rigorous experiments using a haptic tool and experimental setup, we demonstrated their salient features. We demonstrate how ETO and EVO are context-aware and noise-resilient estimation of similarity between the input and output signals. 
% We demonstrated the improved decision making with fine-grained examples, and a long-term average behavior showing up to 62.5$\times$ less time noise for up to \SI{34}{\%} increased RMSE. \todo{this is no longer what we are doing!}
We believe that objective metrics for measuring subjective quality will enable significant improvements in design of  TI applications. While the current work looks at an offline session, we intend to solve the non-trivial problems of measuring objectively in real-time in the near future. 

\section{Acknowledgements}
This project is supported by SCOTT http://www.scottproject.eu that has received funding from the Electronic Component Systems for European Leadership Joint Undertaking under grant agreement No 737422. This joint undertaking receives support from the European Unions Horizon 2020 research and innovation program and Austria, Spain, Finland, Ireland, Sweden, Germany, Poland, Portugal, Netherlands, Belgium, Norway. The authors would like to thank Prof. Eamonn Keogh, UC Riverside, and ir. Luc Hogervorst for providing valuable feedback on the manuscript.
% \todo[inline]{Consider re-adding Eammon Keogh, and to add }
% We demonstrate the effectiveness of the proposed metrics and discuss its potential usage jointly with existing metrics, through examples of informed decision making and a coarse analysis showing 40-150$\times$ less delay adjustments for only \SI{4}{\%}-\SI{17}{\%} increased RMSE.

%% file: 0_main.bbl
\begin{thebibliography}{10}
\providecommand{\url}[1]{#1}
\csname url@samestyle\endcsname
\providecommand{\newblock}{\relax}
\providecommand{\bibinfo}[2]{#2}
\providecommand{\BIBentrySTDinterwordspacing}{\spaceskip=0pt\relax}
\providecommand{\BIBentryALTinterwordstretchfactor}{4}
\providecommand{\BIBentryALTinterwordspacing}{\spaceskip=\fontdimen2\font plus
\BIBentryALTinterwordstretchfactor\fontdimen3\font minus
  \fontdimen4\font\relax}
\providecommand{\BIBforeignlanguage}[2]{{%
\expandafter\ifx\csname l@#1\endcsname\relax
\typeout{** WARNING: IEEEtran.bst: No hyphenation pattern has been}%
\typeout{** loaded for the language `#1'. Using the pattern for}%
\typeout{** the default language instead.}%
\else
\language=\csname l@#1\endcsname
\fi
#2}}
\providecommand{\BIBdecl}{\relax}
\BIBdecl

\bibitem{Fettweis2014}
G.~P. Fettweis, ``The tactile internet: Applications and challenges,''
  \emph{IEEE Vehicular Technology Magazine}, vol.~9, no.~1, pp. 64--70, 2014.

\bibitem{Aijaz2018}
A.~Aijaz and M.~Sooriyabandara, ``The tactile internet for industries: A
  review,'' \emph{Proceedings of the IEEE}, vol. 107, no.~2, pp. 414--435,
  2018.

\bibitem{Millnert2018}
V.~Millnert, J.~Eker, and E.~Bini, ``Achieving predictable and low end-to-end
  latency for a network of smart services,'' in \emph{2018 IEEE Global
  Communications Conference (GLOBECOM)}.\hskip 1em plus 0.5em minus 0.4em\relax
  IEEE, 2018, pp. 1--7.

\bibitem{Samarakoon2018}
S.~Samarakoon, M.~Bennis, W.~Saad, and M.~Debbah, ``Federated learning for
  ultra-reliable low-latency v2v communications,'' in \emph{2018 IEEE Global
  Communications Conference (GLOBECOM)}.\hskip 1em plus 0.5em minus 0.4em\relax
  IEEE, 2018, pp. 1--7.

\bibitem{Gringoli2018}
F.~Gringoli, R.~Klose, M.~Hollick, and N.~Ali, ``Making wi-fi fit for the
  tactile internet: Low-latency wi-fi flooding using concurrent
  transmissions,'' in \emph{2018 IEEE International Conference on
  Communications Workshops (ICC Workshops)}.\hskip 1em plus 0.5em minus
  0.4em\relax IEEE, 2018, pp. 1--6.

\bibitem{holland2019}
O.~Holland, E.~Steinbach, R.~V. Prasad, Q.~Liu, Z.~Dawy, A.~Aijaz, N.~Pappas,
  K.~Chandra, V.~S. Rao, S.~Oteafy \emph{et~al.}, ``The ieee 1918.1 “tactile
  internet” standards working group and its standards,'' \emph{Proceedings of
  the IEEE}, vol. 107, no.~2, pp. 256--279, 2019.

\bibitem{Chaudhari2011}
R.~Chaudhari, E.~Steinbach, and S.~Hirche, ``Towards an objective quality
  evaluation framework for haptic data reduction,'' in \emph{2011 IEEE World
  Haptics Conference}.\hskip 1em plus 0.5em minus 0.4em\relax IEEE, 2011, pp.
  539--544.

\bibitem{Yuan2014}
Z.~Yuan, S.~Chen, G.~Ghinea, and G.-M. Muntean, ``User quality of experience of
  mulsemedia applications,'' \emph{ACM Transactions on Multimedia Computing,
  Communications, and Applications (TOMM)}, vol.~11, no.~1s, p.~15, 2014.

\bibitem{Eid2011}
M.~Eid, J.~Cha, and A.~El~Saddik, ``Admux: An adaptive multiplexer for
  haptic--audio--visual data communication,'' \emph{IEEE Transactions on
  Instrumentation and Measurement}, vol.~60, no.~1, pp. 21--31, 2010.

\bibitem{Cizmeci2017}
\BIBentryALTinterwordspacing
B.~Cizmeci, X.~Xu, R.~Chaudhari, C.~Bachhuber, N.~Alt, and E.~Steinbach, ``A
  multiplexing scheme for multimodal teleoperation,'' \emph{ACM Trans.
  Multimedia Comput. Commun. Appl.}, vol.~13, no.~2, pp. 21:1--21:28, Apr.
  2017. [Online]. Available: \url{http://doi.acm.org/10.1145/3063594}
\BIBentrySTDinterwordspacing

\bibitem{Hinterseer2006}
P.~{Hinterseer}, E.~{Steinbach}, and S.~{Chaudhuri}, ``Perception-based
  compression of haptic data streams using kalman filters,'' in \emph{2006 IEEE
  International Conference on Acoustics Speech and Signal Processing
  Proceedings}, vol.~5, May 2006, pp. V--V.

\bibitem{Gokhale2017}
\BIBentryALTinterwordspacing
V.~Gokhale, J.~Nair, and S.~Chaudhuri, ``Congestion control for network-aware
  telehaptic communication,'' \emph{ACM Trans. Multimedia Comput. Commun.
  Appl.}, vol.~13, no.~2, pp. 17:1--17:26, Mar. 2017. [Online]. Available:
  \url{http://doi.acm.org/10.1145/3052821}
\BIBentrySTDinterwordspacing

\bibitem{Li2018}
\BIBentryALTinterwordspacing
C.~Li, C.-P. Li, K.~Hosseini, S.~B. Lee, J.~Jiang, W.~Chen, G.~Horn, T.~Ji,
  J.~E. Smee, and J.~Li, ``{5G-Based Systems Design For Tactile Internet},''
  \emph{Proceedings of the IEEE}, pp. 1--18, 2018. [Online]. Available:
  \url{https://ieeexplore.ieee.org/document/8452975/}
\BIBentrySTDinterwordspacing

\bibitem{sachs2018}
J.~Sachs, L.~A. Andersson, J.~Ara{\'u}jo, C.~Curescu, J.~Lundsj{\"o}, G.~Rune,
  E.~Steinbach, and G.~Wikstr{\"o}m, ``Adaptive 5g low-latency communication
  for tactile internet services,'' \emph{Proceedings of the IEEE}, vol. 107,
  no.~2, pp. 325--349, 2018.

\bibitem{kim2018}
K.~S. Kim, D.~K. Kim, C.-B. Chae, S.~Choi, Y.-C. Ko, J.~Kim, Y.-G. Lim,
  M.~Yang, S.~Kim, B.~Lim \emph{et~al.}, ``Ultrareliable and low-latency
  communication techniques for tactile internet services,'' \emph{Proceedings
  of the IEEE}, vol. 107, no.~2, pp. 376--393, 2018.

\bibitem{Basdogan2000}
C.~Basdogan, C.-H. Ho, M.~A. Srinivasan, and M.~Slater, ``An experimental study
  on the role of touch in shared virtual environments,'' \emph{ACM Transactions
  on Computer-Human Interaction (TOCHI)}, vol.~7, no.~4, pp. 443--460, 2000.

\bibitem{Sakr2007}
N.~Sakr, N.~Georganas, and J.~Zhao, ``A perceptual quality metric for haptic
  signals,'' in \emph{2007 IEEE International Workshop on Haptic, Audio and
  Visual Environments and Games}.\hskip 1em plus 0.5em minus 0.4em\relax IEEE,
  2007, pp. 27--32.

\bibitem{Salvador2007}
S.~Salvador and P.~Chan, ``Toward accurate dynamic time warping in linear time
  and space,'' \emph{Intelligent Data Analysis}, vol.~11, no.~5, pp. 561--580,
  2007.

\bibitem{Rabiner1975}
L.~R. Rabiner and B.~Gold, ``Theory and application of digital signal
  processing,'' \emph{Englewood Cliffs, NJ, Prentice-Hall, Inc., 1975. 777 p.},
  1975.

\bibitem{Sakoe1978}
H.~Sakoe and S.~Chiba, ``Dynamic programming algorithm optimization for spoken
  word recognition,'' \emph{IEEE Transactions on Acoustics, Speech, and Signal
  Processing}, vol.~26, no.~1, pp. 43--49, 1978.

\bibitem{Chen2005}
L.~Chen, M.~T. {\"O}zsu, and V.~Oria, ``Robust and fast similarity search for
  moving object trajectories,'' in \emph{Proceedings of the 2005 ACM SIGMOD
  international conference on Management of data}.\hskip 1em plus 0.5em minus
  0.4em\relax ACM, 2005, pp. 491--502.

\bibitem{Chen2004}
L.~Chen and R.~Ng, ``On the marriage of lp-norms and edit distance,'' in
  \emph{Proceedings of the Thirtieth international conference on Very large
  data bases-Volume 30}.\hskip 1em plus 0.5em minus 0.4em\relax VLDB Endowment,
  2004, pp. 792--803.

\bibitem{Vlachos2002}
M.~Vlachos, D.~Gunopoulos, and G.~Kollios, ``Discovering similar
  multidimensional trajectories,'' in \emph{icde}.\hskip 1em plus 0.5em minus
  0.4em\relax IEEE, 2002, p. 0673.

\bibitem{berndt1994}
D.~J. Berndt and J.~Clifford, ``Using dynamic time warping to find patterns in
  time series.'' in \emph{KDD workshop}, vol.~10, no.~16.\hskip 1em plus 0.5em
  minus 0.4em\relax Seattle, WA, 1994, pp. 359--370.

\bibitem{Silva2016}
D.~F. Silva and G.~E. Batista, ``Speeding up all-pairwise dynamic time warping
  matrix calculation,'' in \emph{Proceedings of the 2016 SIAM International
  Conference on Data Mining}.\hskip 1em plus 0.5em minus 0.4em\relax SIAM,
  2016, pp. 837--845.

\bibitem{keogh2005}
E.~Keogh and C.~A. Ratanamahatana, ``Exact indexing of dynamic time warping,''
  \emph{Knowledge and information systems}, vol.~7, no.~3, pp. 358--386, 2005.

\bibitem{Bhardwaj2017}
A.~Bhardwaj, B.~Cizmeci, E.~Steinbach, Q.~Liu, M.~Eid, A.~E. Saddik, R.~Kundu,
  X.~Liu, O.~Holland, M.~A. Luden, S.~Oteafy, and V.~Prasad, ``{A Candidate
  Hardware and Software Reference Setup for Kinesthetic Codec
  Standardization},'' in \emph{2017 IEEE International Symposium on Haptic,
  Audio and Visual Environments and Games (HAVE)}, 2017, pp. 53--58.

\bibitem{novintfalcon}
\BIBentryALTinterwordspacing
(2015) Novint falcon - product page. [Online]. Available:
  \url{https://web.archive.org/web/20150215032400/
  http://www.novint.com/index.php/products/novintfalcon}
\BIBentrySTDinterwordspacing

\bibitem{Chai3dfeatures}
\BIBentryALTinterwordspacing
(2019) Chai3d - features. [Online]. Available:
  \url{https://www.chai3d.org/concept/features}
\BIBentrySTDinterwordspacing

\bibitem{netem}
\BIBentryALTinterwordspacing
(2019) networking:netem [wiki]. [Online]. Available:
  \url{https://wiki.linuxfoundation.org/networking/netem}
\BIBentrySTDinterwordspacing

\bibitem{Hasslinger2008}
G.~Ha{\ss}linger and O.~Hohlfeld, ``The gilbert-elliott model for packet loss
  in real time services on the internet,'' in \emph{14th GI/ITG
  Conference-Measurement, Modelling and Evalutation of Computer and
  Communication Systems}.\hskip 1em plus 0.5em minus 0.4em\relax VDE, 2008, pp.
  1--15.

\bibitem{Vittorias2009}
I.~Vittorias, J.~Kammerl, S.~Hirche, and E.~Steinbach, ``Perceptual coding of
  haptic data in time-delayed teleoperation,'' in \emph{World Haptics
  2009-Third Joint EuroHaptics conference and Symposium on Haptic Interfaces
  for Virtual Environment and Teleoperator Systems}.\hskip 1em plus 0.5em minus
  0.4em\relax IEEE, 2009, pp. 208--213.

\bibitem{Ellis2014}
M.~Ellis, D.~P. Pezaros, T.~Kypraios, and C.~Perkins, ``A two-level markov
  model for packet loss in udp/ip-based real-time video applications targeting
  residential users,'' \emph{Computer Networks}, vol.~70, pp. 384--399, 2014.

\end{thebibliography}
